\documentclass[reviewcopy]{elsarticle}

\usepackage{comment}
\usepackage[hidelinks]{hyperref}
\usepackage[]{natbib}
\biboptions{sort&compress}
\hypersetup{colorlinks,urlcolor={blue}}
\journal{the arXiv. Community input is welcome!}
\usepackage{graphicx,array,longtable}
\usepackage[reviewcopy]{adndt}

\begin{document}
\renewcommand{\thefootnote}{\fnsymbol{footnote}}

\begin{frontmatter}

  \title{Critical evaluation of reference charge radii and applications in mirror nuclei}
  
  \author[]{Ben Ohayon}
  \address[*]{The Helen Diller Quantum Center, Department of Physics, Technion–Israel Institute of Technology, Haifa 3200003, Israel, bohayon@technion.ac.il\vspace{-22 pt}}

\begin{abstract}
I present a critical review of absolute root-mean-square charge radii of stable nuclei from $Z=3$ to $Z=32$, which includes a previously overlooked uncertainty in the combined analysis of muonic x-ray and electron scattering experiments. From these \textit{reference radii} and isotope shift measurements, I obtain those of 12 mirror pairs with a traceable and realistic uncertainty budget.
The difference in radii between mirror nuclei is found to be proportional to the isospin asymmetry, confirming recent calculations by Novario \emph{et al}~[PRL~130, 032501]. Assuming that this linear relation holds across the nuclear chart, the fitted proportionality constant, combined with the revised known radii, predicts the radii of 73 previously unknown mirror partners.
These are useful e.g., for benchmarking atomic and nuclear theory, calibrating entire chains, and as an input to nuclear beta-decay calculations.
The radii of $(T=1,T_z=0)$ nuclei are interpolated assuming negligible isospin symmetry breaking.
This completes a model-independent, high-precision extraction of the charge and weak radii of all nuclei involved in the testing of the unitarity of the CKM matrix.
\end{abstract}

\end{frontmatter}
\newpage
\begin{figure}[!h]
\centering
\includegraphics[trim={56 56 28 56},clip,width=0.8\columnwidth]{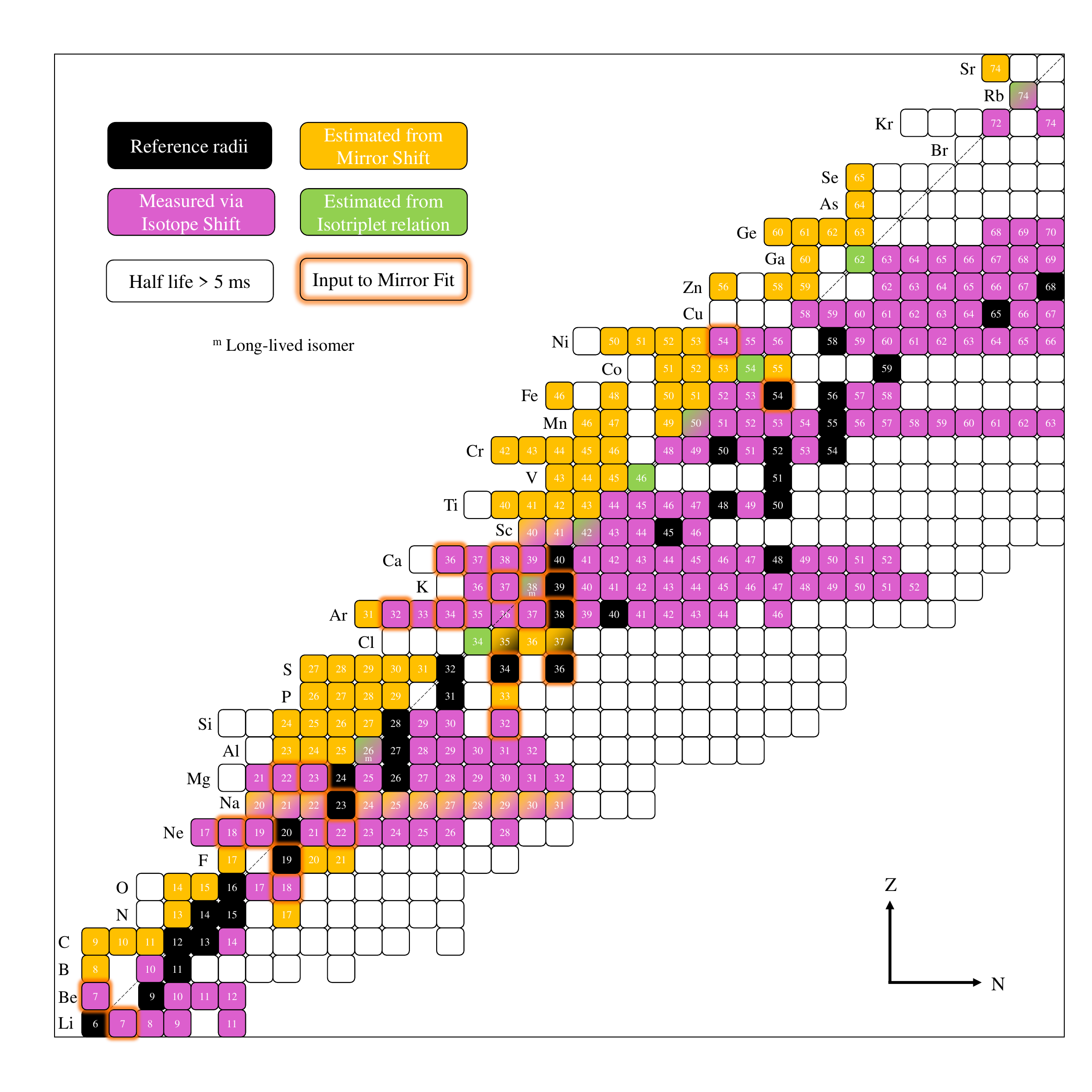}
\caption {
Overview of nuclei relevant to this work, indicated by element name and mass number.
Reevaluated reference radii are given in Tab.~\ref{tab:rad}.
The radii of the neon chain are re-estimated in this work and given in Tab.~\ref{tab:Ne}.
Radii which are an input to the mirror fit are given in Tab.~\ref{tab:fit}.
Radii measured via isotope shifts are given in Tab.~\ref{tab:mirr}, with those
estimated from the mirror shift given in italics.
The recalibrated sodium chain is in Tab.~\ref{tab:Na}.
Charge and weak radii estimated from the isotriplet relation are given in Tab.~\ref{tab:iso}.
}
\label{fig:over}
\end{figure}

\clearpage

\tableofcontents

\section{Introduction}

A key property of atomic nuclei is their finite size, which cannot be precisely calculated from first principles.
It must be measured. 
Knowledge of the size of protons and neutrons and how they distribute within nuclei not only enriches our basic understanding of nuclear structure, but also serves as input to a host of fundamental atomic and nuclear applications. 
%
On the atomic physics side, confronting experiment and theory in light, simple atomic systems require reliable values of root-mean-square (RMS) nuclear charge radii (see e.g.~\cite{2023-Theory} and references therein). 
In heavier simple systems, which are highly charged ions, experiment and theory approach a precision where finite size effects becomes appreciable~(e.g.,~\cite{2019-HCI, 2020-alpha, 2023-gfns, 2023-Tin,2024-U}).
%
In the realm of nuclear physics, refined studies of beta decay are based on accurately determined electroweak moments~\cite{2022-Seng, 2023-EW, 2023-Data, 2025-Al, 2024-K}. 
Particular attention is paid to mirror nuclei, characterized by an interchanged number of protons and neutrons. 
Exploring the radii of mirror nuclei provides insight into isospin effects in nuclei~\cite{2022-Informaiton,2023-ISB}, the extent of the neutron skin~\cite{2023-Mirr}, the equation of state of nucleonic matter~\cite{2021-54Ni}, and the size of neutron stars~\cite{2023-stellar}.

This work serves two purposes, first to provide an up-to-date, transparent, critical evaluation of reference radii in the $Z=3-32$ range where mirror nuclei are relevant.
Second, to predict unknown radii of mirror nuclei, including examples of their applications.
The structure is as follows; in Section~\ref{sec:ref} I reevaluate the radii of the relevant stable nuclei from a combined analysis of muonic atom x-ray spectroscopy and electron scattering experiments. 
In Section~\ref{sec:neon} I update the radii of the neon isotopic chain based on a recent high-precision measurement with coupled ions~\cite{2022-Sailer}.
In Section~\ref{sec:mirror} I update the radii of pairs of mirror nuclei and demonstrate that, as predicted in~\cite{2023-Mirr}, their difference is proportional to the isospin asymmetry. Using the fit results I predict the unknown radii of 73 proton-bound mirror partners.
In Section~\ref{sec:examples} I give two examples of useful applications of the predicted radii to fundamental physics. The first example focuses on the Na isotopic chain, where the predicted radius of $^{21}$Na facilitates a precise benchmark of many-body atomic theory calculations. In the second example, I demonstrate how to precisely predict the charge and weak radii of every nucleus that participates in the determination of the magnitude of the Cabibbo–Kobayashi–Maskawa (CKM) matrix element $V_{ud}$.
A summary is given in Section~\ref{sec:summary}.

\section{Re-evaluation of reference radii}\label{sec:ref}
The backbones of this work are the absolute RMS radii of one or more stable isotope of each element. I refer to them as \textit{reference radii}. 
These have been predominantly obtained from a combined analysis of elastic electron scattering cross sections, and the energies of x-ray photons emitted in the cascade of muons bound to a nucleus~\cite{2004-Fricke} (`muonic x-rays').
The procedure used in most cases is described in e.g.~\cite{2022-review}. 
First, a Barret equivalent radius (BER), $R^\mu_{k,\alpha}$, is extracted from a measured transition energy. The superscript $\mu$ is there to remind us muonic atoms are used.
The parameters $k\approx2,\alpha\approx0$ are calculated for that transition (typically $2P_{3/2}-1S$) by numerically solving the Dirac equation with some approximate charge distribution~\cite{1970-Barrett}. They can be found in~\cite{2004-Fricke}.
In the spherically symmetric limit, $R^\mu_{k,\alpha}$ is by construction nuclear charge distribution independent~\cite{1995-Fricke}.
In all but a few light cases~\cite{2023-QUARTET}, the accuracy of the measured energies is high, so the uncertainty in $R^\mu_{k,\alpha}$ is dominated by that of the nuclear polarization correction to the muonic energy levels. This quantity is typically an intricate combination atomic and nuclear theory calcinations with possibly some experimental input [].
 
In order to translate $R^\mu_{k,\alpha}$ to the more universal RMS charge radius, information on the charge distribution (`nuclear shape'), from either calculations or elastic electron scattering experiments, is incorporated.
It has been noticed that such experiments are particularly well-suited to measure \textit{ratios} of moments, typically with accuracies an order of magnitude higher than integral quantities~\cite{1995-Fricke}.
Denoting the ratio $V_2^e=R^e_{k,\alpha}/r^e\approx\sqrt{5/3}$, with the superscripts to remind us that these quantities are calculated from charge distributions measured in electron scattering, then the RMS radius obtained by a combined analysis reads
\begin{equation}\label{eq:combined}
r^{e\mu}=R^\mu_{k,\alpha}/V_2^e.
\end{equation}

$r^{e\mu}$ is considered to be highly reliable as it exploits the strength in the two experimental methods while circumventing their weaknesses~\cite{1995-Fricke}.
However, there are certain limitations to this recipe. The main one is that high-statistics, high momentum-transfer scattering experiments, as well as the individuals who interpret them, are rare, making the applicability of Eq.~(\ref{eq:combined}) limited to a subset of the cases.
To circumvent this problem, Angeli had fitted an analytical function to a subset of the $V_2$ factors, averaged on isotopes, making a coarse estimate of the variation $V_2(Z)$~\cite{1998-Ang}. The resulting $V_2$ factors are only accurate to a few per mill (as stated in Ref.~\cite{1998-Ang}), which translates directly to a similar fractional accuracy in charge radii determined via a combined analysis of Eq.~(\ref{eq:combined}). These considerations are absent in later compilations~\cite{2004-Ang, 2013-AM}, in which no uncertainty in $V_2(Z)$ is given.
To my knowledge, Ref.~\cite{2021-Al}, in which $r^{e\mu}$($^{27}$Al) is estimated, is the only example in which an uncertainty in $V_2$ is given. 
To accomplish this, $V_2$($^{27}$Al) was estimated from three different scattering experiments~\cite{1967-AlSc,1973-AlScat, 1987-VJV}. The difference between the results is taken as uncertainty, which was found to dominate that of $r^{e\mu}$($^{27}$Al).
Here I refine and generalize this approach in order to recommend transparent  reference radii which include an uncertainty due to the nuclear distribution shape.
\begin{figure}[!tbp]
\centering
\includegraphics[trim={70 128 100 216},clip,width=0.62\columnwidth]{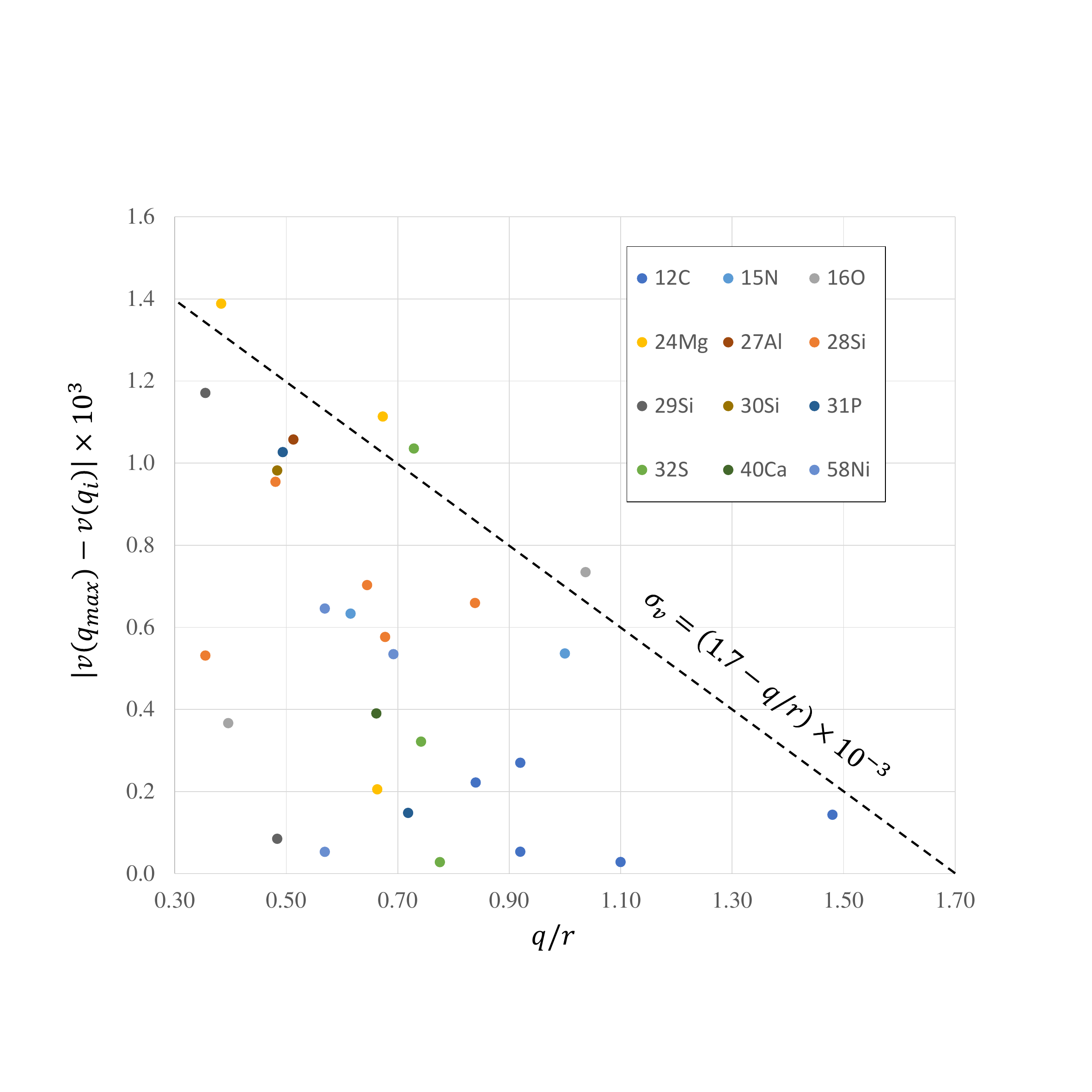}
\caption {
Estimation of uncertainty in $v$ factors based on a comparison of results at lower-than-maximal momentum transfer.
}
\label{fig:sigmav}
\end{figure}

In light of the above, I now show that the uncertainty in $V_2$ is dominating for most radii and must be taken into account. 
For convenience, I introduce the reduced proportionality factor
\begin{equation}\label{eq:nu}
    v=1-\sqrt{\frac{3}{5}}V_2^e,
\end{equation}
which tends to zero for nuclei with a uniform charge distribution.
We will see that $v$ increases from 2 to 9 per mill between $Z=6$ and $Z=32$. This means that any extraction of charge radii from muonic x-ray measurements with sub percent accuracy must include a discussion of the nuclear shape.
Following Ref.~\cite{2015-qr}, I note that in scattering experiments, a high momentum transfer (measured e.g. in fm$^{-1}$) compared to the nuclear size is needed in order to be sensitive to higher moments of the charge distribution. Thus, the maximal momentum transfer over the RMS radius, $q/r$, is indicative of the ability of a scattering experiment to resolve subtle features of the charge distribution.
For each nucleus on which electrons were scattered upon, I estimate the most probable value of $v$ from the charge distribution obtained by the scattering experiments with the highest maximal $q$.
These data are usually the ones in which a (nearly) model-independant analysis of the scattering data was done. 
The results are given in Tab.~\ref{tab:v}.

\begin{figure}[!tbp]
\centering
\includegraphics[trim={8 182 30 200},clip,width=0.62\columnwidth]{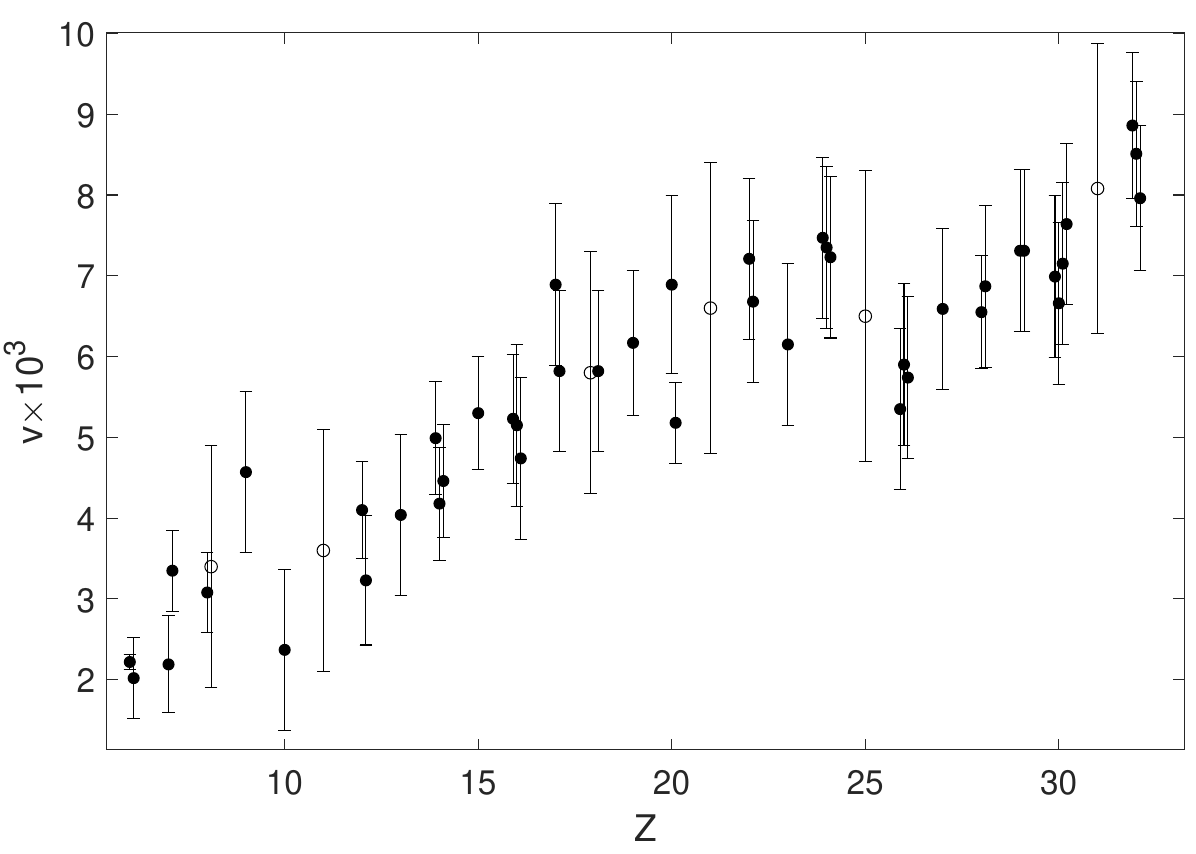}
\caption {
Recommended $v$ factors, with uncertainty estimated using Eq.~(\ref{eq:unc}).
Open symbols denote interpolated values.
Small horizontal shifts were introduced for clarity.
}
\label{fig:nu}
\end{figure}

Having chosen the most probable value, $v(q_\mathrm{max})$, I turn to bootstrap its uncertainty. 
To do that, for each nucleus in which one or more measurements with a high momentum scattering are available, I extract the less accurate factors $v(q_i)$'s from the rest of the scattering experiments whose parametrizations are given in Ref.~\cite{1987-VJV, 1995-Fricke, 1997-WE}. 
I then look at the deviation $|v(q_i)-v(q_\mathrm{max})|$, shown in Fig.~\ref{fig:sigmav}.
It is clearly seen that the lower the figure of merit $q/r$, the higher the chance the the $v$ factors will significantly deviate from their most probable values.
In several cases, the deviation is unreasonably low, which I ascribe to correlations in their analysis. These are difficult disentangle due to the partiality of the published information. In light of this, I parameterize the uncertainty taking into account the most deviating points.
To account for these possible deviations, I choose a simple uncertainty parametrization
\begin{equation}\label{eq:unc}
    \sigma_v(q,r)=(1.7-q/r)\times10^{-3},
\end{equation}
as shown in Fig.~\ref{fig:sigmav}.
The most sensitive second-best experiment, was in $^{12}$C, reaching $q=4.0\,$fm$^{-1}$ so that $q/r=1.6~$fm$^{-2}$ and so, according to Eq.~(\ref{eq:unc}), $\sigma_v=10^{-4}$. However, the majority of experiments are in the range $q/r=0.5-0.9~$fm$^{-2}$ resulting in $\sigma_v\sim10^{-3}$.
For some nuclei relevant to this work no reliable scattering experiments are available. I estimate their $v$ factors by interpolating from the neighboring elements as shown in Fig.~\ref{fig:nu}. Generous uncertainties are given for these interpolated values.

\begin{figure}[tbp]
\centering
\includegraphics[trim={8 10 20 10},clip,width=0.86\columnwidth]{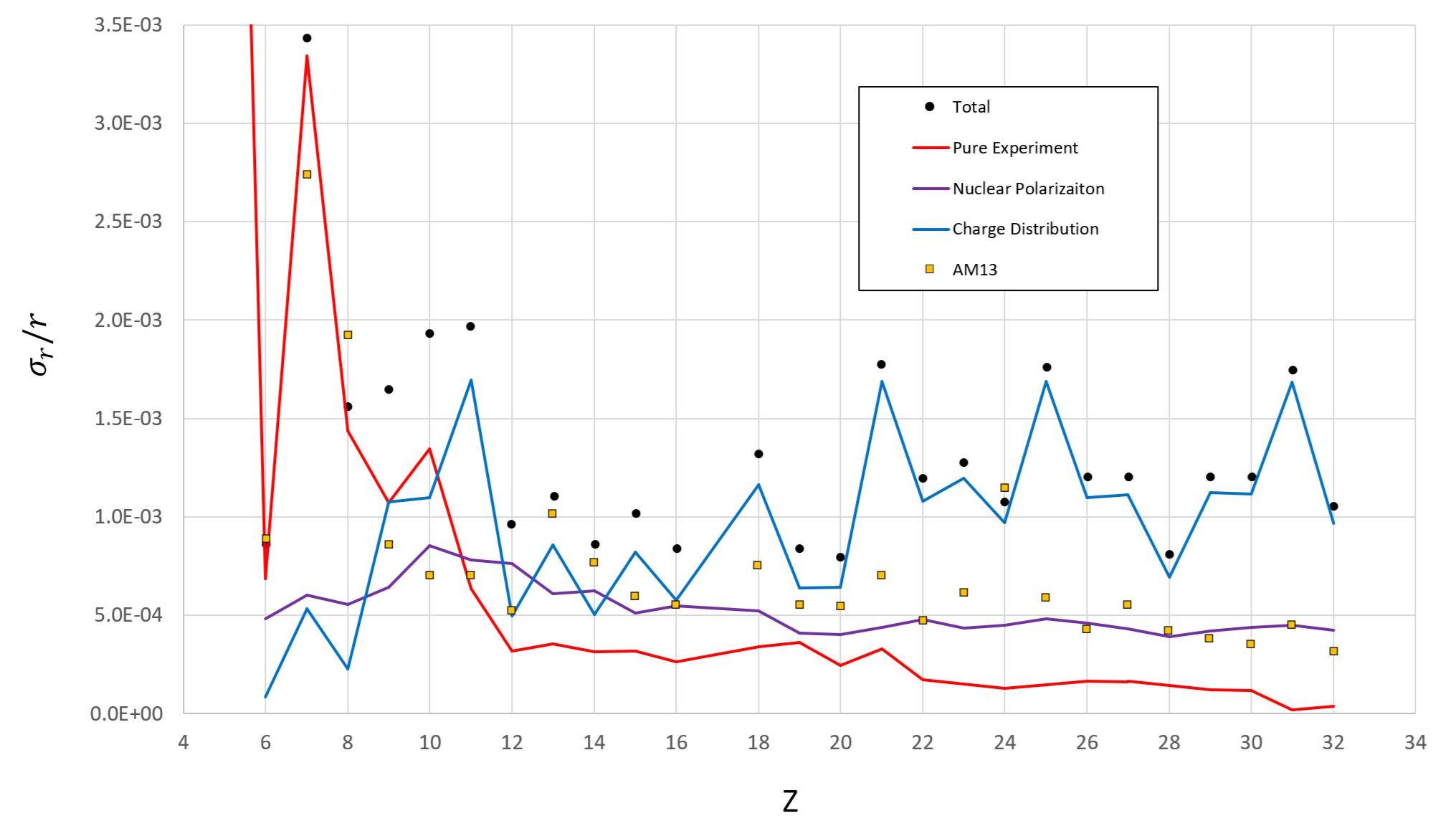}
\caption {
Contributions to the uncertainty of reference radii as determined from muonic x-ray measurements. 
For each element, the numbers correspond to the isotope whose radius is best known.
The square markers denote the corresponding total uncertainties given in~\cite{2013-AM}.
}
\label{fig:unc}
\end{figure}

Employing Eq.~(\ref{eq:nu}), the $v$ factors given in Tab.~\ref{tab:v}, and the BERs collected in~\cite{2004-Fricke}, I reevaluate the reference radii given in Tab.~\ref{tab:rad}. 
For the light nuclei $^6$Li, $^9$Be, $^{11}$B, $^{15}$N, as well as the medium mass Cl isotopes, radii are not sufficiently well measured by muonic x-ray spectroscopy. I thus use values from scattering experiments (see also discussion in Ref.~\cite{2023-QUARTET}).
For each nucleus whose radius is extracted from muonic x-ray experiments, three distinct uncertainty contributions are taken into account. The first is related to the `pure' experimental one, which stems from the statistical precision and calibration accuracy in the measured energies of the relevant transitions.
The second is related to the uncertainty in the nuclear polarization correction as given in~\cite{2004-Fricke}.
The third uncertainty is directly from that in the $v$ factors.
The different fractional uncertainty contributions are plotted in Fig.~\ref{fig:unc}. 
From the figure we readily see that from carbon to neon, experimental uncertainties dominate. From phosphorous onward, it is the charge-distribution uncertainty, unaccounted-for in prior compilations,  is dominant.
The trend continues, so that even larger missing uncertainties are expected for the heavier nuclei, as was recently pointed out~\cite{2023-deform}.

I also plot in Fig.~\ref{fig:unc} the total fractional uncertainties given in the latest compilation~\cite{2013-AM}.
As expected, they roughly follow the nuclear polarization trend above $Z=10$, thus emphasizing the need for accounting for $\sigma_\mathrm{CD}$, which are up to 4 times larger.
Adopting the values in~\cite{2013-AM} would result in shifts of on average one of their reported uncertainties. this illustrates that the impact of the choice of the calibration factors in Tab.~\ref{tab:v}, and demonstrates the importance of the increased uncertainty budget. 

Having reevaluated the reference radii, I combine them with isotopic differences $\delta r$ (or $\delta r^2$). These differences are mostly extracted from isotope shift measurements (reviewed e.g. in~\cite{2023-ExpRev}) combined with theory calculations (see e.g.~\cite{2024-review}).
The subset of radii relevant to mirror nuclei is given in Tab.~\ref{tab:mirr}, in which references to the isotope shift measurements and atomic theory literature can be found.
For a small set of the mirror nuclei, both the radius of the proton-rich and of the neutron-rich partner has been measured.
Their difference is used in the following to make the mirror fit. Before that, I digress to improve the extraction of the radii of the neon chain, which plays a key role in the fit.

\section{Re-calibration of the neon chain}\label{sec:neon}

Being a light element in which optical isotope shifts for several proton-rich isotopes were measured~\cite{2008-Ne,2011-Ne}, the neon chain plays a key role in this work. I now show that incorporating a recent novel measurement of the bound-electron g-factor isotope shift in hydrogen-like neon significantly improves the radii of the entire chain, which harbors important proton-rich isotopes taking part in the mirror fit.

First, I comment on the reference radius of the $^{20}$Ne nucleus. It is given in Tab.~\ref{tab:rad} as $r_{20}=3.001(6)\,$fm. Its uncertainty is mostly from experiment, and with equal contributions from nuclear polarization, and charge distribution parametrization.
It is also in agreement with $r_{20}=3.006(5)$~fm from Ref.~\cite{2004-NeFrick},  which is estimated from the corresponding BER using a two-parameter Fermi distribution with surface thickness $t=2.3\,$fm. I estimated the uncertainty by changing $t$ by $10\%$, while keeping $r_{20}\sim3.0\,$fm and recording the variation in $v$.
Another radius can be found in a recent compilation $r_{20}=3.0055(21)\,$fm~\cite{2013-AM}, with uncertainty smaller than the quadratic sum of experimental and nuclear polarization uncertainties.
It is worth noting that I estimate a smaller radius than the others. The reason for this is that I calculated the calibration factor $v=2.4(1.1)$ from the three Fermi distribution parameters given in~\cite{1987-VJV} (Denoted by~``Be85"), due to the relatively high $q_{eff}=1.8\,$fm$^{-1}$ range covered. 
Adopting the charge distributions from either \cite{1981-NeScat} or \cite{1971-NeScat}, which only extend to $q_{eff}=1\,$fm$^{-1}$, results in $v\sim5$ giving $r_{20}=3.009\,$fm.
The $0.3\%$ difference in the radius obtained using $v$-factors from different scattering experiments is a testament to the importance of accounting for uncertainty in the charge distribution.
The lack of high momentum transfer scattering data in $^{22}$Ne prevents us from estimating its radius directly from its Barrett radius given in~\cite{2004-NeFrick}.
Assuming $v_{22}=v_{20}\pm0.5$, in line with the isotopic variations from Tab.~\ref{tab:v}, returns a difference 
\begin{equation}\label{eq:drNe}
  \delta r^2_{20,22}=-0.310(16)_\mathrm{exp}(5)_\mathrm{NP}(9)_\mathrm{v}\,\mathrm{fm}^2,  
\end{equation}
where the uncertainty from a possible variation of the charge distribution between the isotopes is not negligible.

Optical isotope shifts were measured for the $614\,$nm transition in a long chain of neon isotopes~\cite{2008-Ne,2011-Ne}.
To calibrate the radii of the neon chain, a partial king plot procedure is used. It relies on the IS equation
\begin{equation}
\delta v^{A,A'}_i \approx
K_i\mu^{A,A'} + F_i  (\delta r^2)^{A,A'},
\label{eq:IS}
\end{equation}
with $F_{614}$ calculated via many-body atomic theory and one differential radii pair to determine $K_{614}$. The resulting $K_{614}$ and calculated $F_{614}$ are then used to extract $\delta r^2$ from optical isotope shifts.
Originally, $F_{614}=-40(4)\,$MHz/fm$^2$ was estimated semi-empirically using the Goudsmit-Fermi-Segre method (GFS)\,\cite{2008-Ne}. Resulting in the radii of the chain that were limited by $\sigma_F$. A later \textit{ab initio} calculation returned $F_{614}=-30.5(1.5)$MHz/fm$^2$\,\cite{2019-Ne}, where I adopt here the more conservative uncertainty estimate given in~\cite{2021-NeF}. 
The disagreement between the semi-empirical and \textit{ab initio} methods is attributed to the limited accuracy of the GFS formula, as discussed e.g. in~\cite{1985-FS, 1992-Ca, 2006-Mg, 2019-Ne}.
Using this $F_{614}$, the radii of the chain could be improved by up to a factor of $1.6$~\cite{2019-Ne}, with their uncertainty dominated by that of $\delta r^2_{20,22}$ given in Eq.~(\ref{eq:drNe}).
Thus, it is clear that a better determination of $\delta r^2_{20,22}$ would increase accuracy far from stability. 

Fortunately, a new method has recently emerged to determine differential radii in even-even isotopes. This is accomplished through measuring the differences in the g-factors of single electron bounds to bare nuclei.
For neon, such a measurement returned $r_{20}-r_{22}=-0.0533(4)\,$fm corresponding to $\delta r^2_{20,22}=-0.3171(24)\,$fm$^2$~\cite{2022-Sailer}. A notable improvement by factor $8$ over the value given in Eq.~(\ref{eq:drNe}).
Applying Eq.~(\ref{eq:IS}) with the highly accurate $\delta r^2_{20,22}$ from the g-factor measurement results in improved differential radii of the entire neon chain. They are given in Tab.~\ref{tab:Ne}. These are more precise by up to a factor $2.5$ as compared with~\cite{2019-Ne}. The improvement was somewhat curtailed by my more conservative uncertainty estimation for $r_{20}$ and $F_{614}$.
Notably, the radius of the superallowed beta-emitter $^{18}$Ne, is shifted by $5\sigma$ compared with the value quoted in~\cite{2013-AM}.
The current uncertainty budget motivates both new optical measurements with higher precision and a more accurate calculation of $F_{614}$.
%
This example shows the tremendous impact of measuring a single $\delta g_\mathrm{bound}$, and motivates extending such measurements to other stable even-even pairs.

\section{The mirror shift fit}\label{sec:mirror}
\begin{figure}[tbp]
\centering
\includegraphics[trim={26 180 50 200},clip,width=0.66\columnwidth]{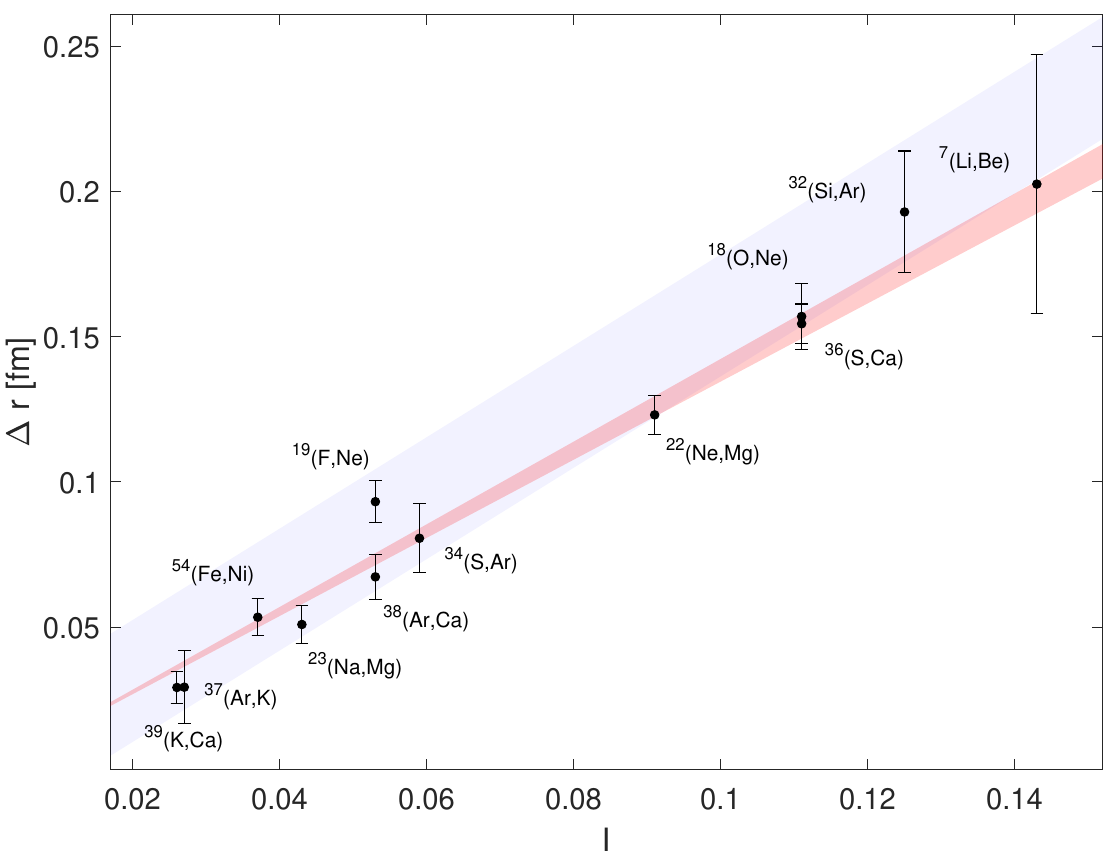}
\caption {
Linear fit to the mirror shift. Data-points are the individual shifts from Tab.~\ref{tab:fit}. The dark shaded region is the $68\%$ confidence interval of the fitted slope given in Eq.~(\ref{eq:fit}). The light shaded region is the $68\%$ confidence interval calculated from first principles~\cite{2023-Mirr}.
}
\label{fig:mirr}
\end{figure}

Extensive \textit{ab initio} calculations suggest that the differences in radii between mirror nuclei (mirror shifts) are proportional to the isospin asymmetry $I=(N-Z)/A$, at least up to $I\approx0.2$~\cite{2023-Mirr}.
The experimental situation is not as healthy.
In contrast with isotope shifts, which can be measured directly via optical spectroscopy, mirror shifts are difficult to measure with high fractional precision. To obtain them, one has to take the difference between the absolute radii of the mirror pair, contending with large experimental and theoretical uncertainties, which are emphasized in the previous sections.
Nevertheless, with the reference radii given in Tab.~\ref{tab:rad}, $\delta r^2$ from the literature (the references are given in Tab.~\ref{tab:mirr}), and the re-calibrated neon differential radii given in Tab.~\ref{tab:Ne}, we have all the ingredients to test the linear theoretical prediction experimentally.
The relevant data is given in Tab.~\ref{tab:fit}. I excluded the pair $^{21}$Na-$^{21}$Ne, as its uncertainty is too large to be of use. The pairs with Cl isotopes were not included as their radii were determined from electron scattering experiments, the interpretation of which depends on the nuclear model.
The result of a simple one-parameter fit to the weighted mirror shifts is shown in Figure~\ref{fig:mirr}. It has a reduced Chi-square of $11.5/11=1.04$ indicating that such a fit is not inconsistent with the data.
The resulting mirror shift parametrization is
\begin{equation}\label{eq:fit}
    \Delta_I=r_{N,Z}(I)-r_{Z,N}(I)=1.382(34)\times I ~\mathrm{fm}.
\end{equation}
Almost half of the uncertainty given in Eq.~(\ref{eq:fit}) stems from that of the charge distributions, previously overlooked in the literature. Omitting it would result in a reduced $\chi^2$ of $2.3$. The second-largest contributor is the uncertainty in the differential radii, as extracted from optical isotope shift measurements. It originates mostly from the atomic factors and not the statistical accuracy of the optical measurements (see e.g.~\cite{2012-Role}).
The empirically determined slope of Eq.~(\ref{eq:fit}) agrees with the band spanning nuclear theory calculations~$\Delta_I=(1.574\times I)\pm0.021~$fm~\cite{2023-Mirr}, also shown in Figure~\ref{fig:mirr}.
To extend the comparison between experiment and theory, it would therefore be very interesting to add experimental data at high asymmetry $0.12<I$ for which precise data are currently not available. 
The pair which weight is the largest in the fit is $^{36}$Ca$-^{36}$S, which combines high experimental accuracy with a large asymmetry.
The pair that sits farthest from the fit, $2.3$ standard errors away, is $^{19}$F$-^{19}$Ne. Excluding it reduces the reduced $\chi^2$ to $0.59$. It could be due to a statistical deviation, an underestimated error in the radii determinations, or a genuine nonlinearity.

Having validated that the mirror shift is indeed proportional to $I$, I can use the fit results of Eq.~(\ref{eq:fit}) to predict the unknown radii for mirror partners of nuclei which radii are known (colored orange in Fig.~\ref{fig:over}). 
The results are given in italics in Tab.~\ref{tab:mirr}, where the radii of $73$ mirror partners are predicted. It is noted that I only focus on proton-bound nuclei~(see e.g.~\cite{2025-MirrorMAss}).
The uncertainties in the predicted radii are mostly dominated by that of the measured partner and not by the uncertainty in the mirror shift fit. This means that small deviations from linearity, such as that of the $^{19}$F-$^{19}$Ne pair, have a mostly negligible effect on my predicted radii.

\section{Example applications of the predicted radii}\label{sec:examples}

\subsection{Testing atomic many-body calculations}

Optical isotope shifts have been measured for a long chain of Na isotopes~\cite{1978-NaIS}. However, extracting $\delta r^2$ from these measurements has been challenging. This is because Na is both mono-isotopic, preventing muonic x-ray measurement in more than the stable isotope, and light - making the needed accuracy in calculating its mass-shift factor very stringent.
Specifically, a recent calculation of the IS factors of Na, combined with isotope shift measurements, returned $r_{21}=2.97(1)[6]\,$fm with uncertainty dominated by that of the calculation of the mass shift factor~\cite{2022-NaMg}. This value is not accurate enough to be useful in the mirror fit. However, we can inverse this logic and use Eq.~(\ref{eq:fit}) to get $r_{21}=3.029(7)\,$fm which agrees with the literature value but enjoys an uncertainty smaller by an order of magnitude. Combining the reference radius $r_{23}=2.992(6)\,$fm from Tab.~\ref{tab:rad} returns the MS difference $\delta r^2_{23,21}=0.22(5)\,$fm$^2$. It is much more accurate, and in agreement with $\delta r^2_{23,21}=-0.13(5)[34]\,$fm$^2$ given in~\cite{2022-NaMg}.

Armed with one MS radius difference in the Na chain, I can now calibrate its entirety using the calculated FS factor, which is considered reliable~\cite{2022-NaMg}.
Employing Eq.~(\ref{eq:IS}) with $\delta r^2_{23,21}=0.22(5)\,$fm$^2$ and $F=-39.3(3)\,$MHz/fm$^2$ returns $K_{D1}=384.7(6)\,$GHz\,u.
This MS factor agrees with, and is more accurate than, that which is calculated with relativistic coupled cluster method $K_{D1}=388(3)\,$GHz\,u~\cite{2022-NaMg}. It thus sets a benchmark for the next generation calculations~\cite{2024-review}.
The mass shift also agrees with a semi-empirical estimation based on a combination of matter radii and calculated neutron skins $K_{D1}=384.0(1.3)$~\cite{2022-NaMg}.
Half of the uncertainty in $K_{D1}$ is from $\delta r^2_{23,21}$, in turn stemming from the absolute radii of $^{20}$Ne and $^{23}$Na, and the other half is from the $D_1$ line optical isotope shift measurement in $^{21,23}$Na.
A more precise measurement of this isotope shift would increase the accuracy of $K_{D1}$ and so the radii of the entire Na chain. 
The extracted radii of the chain based on these values, along with their mirror partner values, are given in Tab.~\ref{tab:mirr}.

\subsection{Charge and weak radii of CKM-determining isotopes}

The largest CKM matrix element $|V_{ud}|$ is extracted from measured parameters (half-lives, Q-values, etc.) in superallowed $0^+$$\rightarrow0^+$ beta transitions between states with isospin $T=1$.
It is usually parameterized as
\begin{equation}
    V_{ud}^{-2}\propto ft(1+\delta_R')(1+\delta_\mathrm{NS}-\delta_C)(1+\Delta_R^V),
\end{equation}
with $f(Q)$ the statistical rate function, $t$ the beta decay half-life, $\Delta_R^V$ is a nucleus-independant (universal) correction, and $\delta_i$ are various nucleus-dependent corrections; their uncertainty now dominates that of the
currently accepted value of $|V_{ud}|$~\cite{2020-TH}.

\begin{figure}[tbp]
\centering
\includegraphics[trim={10 232 32 246},clip,width=0.7\columnwidth]{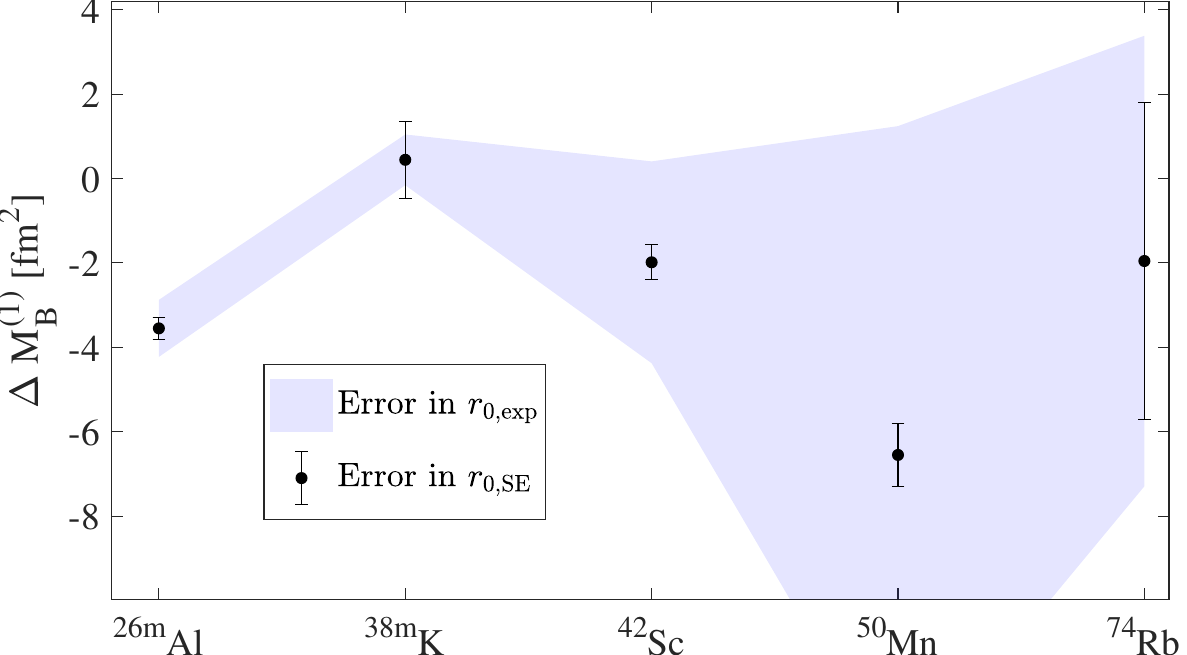}
\caption {
Testing for isospin symmetry breaking by comparing measured (exp) and semi-empirical (SE) radii. See Eq.~(\ref{eq:DeltaM}) and Tab.~\ref{tab:iso}.
}
\label{fig:T0}
\end{figure}

Recently, the role of nuclear charge radii in the calculation of $f$ has been put into the spotlight~\cite{2023-Data, 2025-Al}, pointing out that the effect of their uncertainty is much larger than previously considered. 
Moreover, it has been recognized that absolute radii may constrain the isospin symmetry breaking correction $\delta_C$ as well~\cite{2023-EW, 2024-K}.
Here, the empirical mirror relation already gives rise to reliable estimation of the radii of all nuclei with $T_z=-1$ that are involved in the determination of $|V_{ud}|$ (see Tab.~\ref{tab:mirr}).
However, some $T_z=0$ nuclei play a key role as well~\cite{2020-TH}, with only a handful of their radii measured. 
To estimate the radii of these nuclei, I first denote them by $r_{Tz}$ with $T_z=-1,0,+1$. 
The mirror fit then translates to
\begin{equation}\label{eq:Delta}
r_{-1}^2-r_{+1}^2=\Delta_I (2r_{\pm1}\pm\Delta_I)    
\end{equation}
 with $\Delta_I$ given in Eq.~(\ref{eq:fit}).
This form is suitable for combining with equation~16 from \cite{2022-Seng}, to obtain a semiempirical (denoted SE) isotriplet interpolation formula for the radius of $T_z=0$ nuclei
\begin{equation}\label{eq:r0}
    r_{0,\mathrm{SE}}^2=r_{\pm1}^2\pm\frac{Z_{\mp1}}{2 Z_0} \Delta_I (2r_{\pm1}\pm\Delta_I).
\end{equation}
The radii of $T_z=0$ nuclei estimated with Eq.~(\ref{eq:r0}) are given in Tab.~\ref{tab:iso}. Their uncertainties span $0.1-1.5\%$ and are dominated by that of $r_{+1}$.
The least well-known triplet is that with $A=10$, motivating an improved determination of the radius of $^{10}$Be.

If all else is under control, and spin-orbit corrections within a triplet are neglected, then the difference between experimental and semi-empirical radii can help to search for, or constrain, isospin-symmetry-breaking (ISB) within the isotriplets.
Plugging Eq.~(\ref{eq:r0}) to Eq.~(10) from Ref.~\cite{2023-EW} I obtain the compact expression
\begin{equation}\label{eq:DeltaM}
    \Delta M_B^{(1)}\approx Z_0(r_{0,\mathrm{SE}}^2-r_{0,\mathrm{exp}}^2),
\end{equation}
which vanishes in the isospin-symmetric limit. Eq.~(\ref{eq:Delta}) provides a simple and intuitive relation between the differences of measured and interpolated radii of $T_z=0$ nuclei and $\Delta M_B^{(1)}$.
The results are given in Tab.~\ref{tab:iso}, and plotted in Fig.~\ref{fig:T0}.
The most stringent constraint on ISB is with the $A=38$ triplet, for which $|\Delta M_B^{(1)}(38)|\leq1.5\,$fm$^2$, comparing well with theoretical predictions, which are as high as $|\Delta M_B^{(1)}(38,\mathrm{DFT})|\leq0.42\,$fm$^2$~\cite{2023-EW}. 
For the $A=26$ triplet, one may conclude that ISB effects contribute at the level of 5 standard deviations $\Delta M_B^{(1)}(26)=-3.5(0.7)\,$fm$^2$. However, the magnitude of the largest calculated effect is much smaller, at $\Delta M_B^{(1)}(26,\mathrm{DFT})=-0.12\,$fm$^2$~\cite{2023-EW}.
Further scrutiny is advised before drawing any conclusion about the existence of large ISB corrections: 1. Measurement of isotope shifts in more transitions and testing of the same MS radius difference is extracted from each~(see e.g.~\cite{2024-Ag}). 2. Perform more independent calculations with different methods~\cite{2024-review}).
3. Measuring $r$($^{26}$Al) directly with muonic atom x-ray spectroscopy of a microgram-scale target~\cite{2023-RefRad}, and then extract $r$($^{26m}$Al) from the isomer shift which is largely independent of the most difficult part of the many-body atomic theory calculation.

I now turn to estimating the weak radius of the triplets, which is of interest, e.g., for calculating recoil corrections involved in extracting the $|V_{ud}|$ matrix element~\cite{2022-Seng}. Combining Eq.~(\ref{eq:Delta}) and Eq.~(16) in~\cite{2022-Seng} gives another simple formula
\begin{equation}\label{eq:rwk}
    r_\mathrm{CW}^2=r_{\pm1}^2+\frac{Z_{\mp1}}{2} \Delta_I (2r_{\pm1}\pm\Delta_I),
\end{equation}
which connects weak radii with the well measured ones of $r_{+1}$ and the result of the mirror fit $\Delta_I$. I note that the role of $\Delta_I$ in Eq.~(\ref{eq:rwk}) is larger, by a factor $Z_0$ than its role in Eq.~(\ref{eq:r0}).
Consequently, the weak radii calculated via Eq.~(\ref{eq:rwk}) and given in Tab.~\ref{tab:iso}, have an uncertainty dominated primarily by that of $\Delta_I$, and not $r_{+1}$.
This means that the key to improving $r_\mathrm{CW}^2$ for most isotriplets is to reduce the error in the mirror slope $\Delta_I$.
The notable exception is the $A=10$ triplet, which would also benefit from a more accurate measurement of $r(^{10}$Be).
I also quote in Tab.~\ref{tab:iso} the $R_\mathrm{CW}^2$ from Ref.~\cite{2022-Seng}, only available for half of the cases of interest.
For the known cases, I reduced the uncertainty by, on average, a factor of 13.
A notable difference in $R_\mathrm{CW}^2$, by the $2.4$ standard deviations, is observed for $A=18$. It originates from the missing systematic uncertainties in the tabulation of Ref.~\cite{2013-AM}, and from the shifted field shift factor as discussed in Section~\ref{sec:neon}.
Our correction to the neon chain removes the large deviations from linearity seen in Fig.~5 of Ref.~\cite{2024-Bayes}. 

\section{Summary}\label{sec:summary}

In this work I reanalyzed the absolute root-mean-square charge radii of stable nuclei up to $Z=32$ using the conventional `Barret moment' recipe for combined analysis of muonic x rays and electron scattering. 
I showed that even within this simplified framework, there is a notably large missing uncertainty contribution to the reference radii from my knowledge of the charge distribution. 
A shortcoming of this method which is emphasized by this work is that the Barret recipe prevents us from benefiting from possible cancellations between the elastic and inelastic part of the two-photon exchange, as is seen in light systems where a perturbative expansion is suitable~\cite{2020-LiNS}.

The reference radii were combined with isotopic differences, stemming from measured electronic isotope shifts and atomic theory calculations, to obtain those of isotopic chains. 
In some cases, large uncertainties in the theory have not been taken into account, which is remedied here.
More emphasis is placed on the neon chain, which was calibrated using a novel measurement in coupled high-charged ions~\cite{2022-Sailer}.
The physics discussed in this work would greatly benefit from extending such measurements to more even-even pairs, notably of O, Si, Ar, and Ca.

Having constructed the database, I focus on 12 pairs of mirror nuclei with reasonably measured radii. Their difference is found to be proportional to the isospin asymmetry, which confirms the calculations of~\cite{2023-Mirr}, at least for the nuclei with $18<A$ and $I<0.12$. From the fitted proportionality constant and its uncertainty, the radii of 73 previously unknown mirror partners are predicted.

Two examples of applications benefiting from these new and or improved radii are given. First, I use the predicted radius of $^{21}$Na with the calculated field shift factor of a relevant transition to determine its mass shift factor. This factor is used to recalibrate the radii of the chain spanning $^{20}$Na to $^{31}$Na, also giving the radii of their mirror partners, such as $^{20}$F. The factor also compares favorably with atomic many-body calculations~\cite{2022-NaMg}, setting a stringent benchmark for the next generation of calculation methods~\cite{2024-review}.

The second example deals with radii of isotriplet nuclei as sandbox for isospin symmetry breaking studies.
Assuming that symmetry breaking is negligible as compared with the relevant uncertainties. In this case, I can combine the fit result with the relation from~\cite{2022-Seng} to obtain a simple equation for determining unknown radii of $T=1,T_z=0$ nuclei from the mostly well known radii of those with $T=1,T_z=1$.
This completes a model-independent, high-precision extraction of the charge radii of all nuclei involved in the testing of the unitarity of the CKM matrix.
Relaxing the assumption of zero isospin symmetry breaking effects enables to test it. Without the mirror fit, this could only be done with the $A=38$ triplet. Here, four more cases are considered. I find a general agreement except for with the $A=26$ case. This calls for refined isotope shift measurements in the Aluminum chain, more atomic factor calculations of the relevant transitions, and a direct measurement of the radioactive $^{26}$Al radius with muonic atoms and microgram targets.
Extending recent measurements from neutron- to proton-rich Si would be beneficial as well.
Finally, I extract the weak radii of all nuclei involved in the test of the unitarity of the CKM matrix. Half of these weak radii were previously unknown, and the rest have improved by an order of magnitude.

\section*{Acknowledgement}
I thank CY~Seng, M~Gorshteyn, M~Deseyn, and M~Heines for their insightful comments.
The support of the Council for Higher Education Program for Hiring Outstanding Faculty Members in Quantum Science and Technology is acknowledged.

\datatables 
\LTright=0pt
\LTleft=0pt

\normalsize
\begin{longtable}{@{\extracolsep\fill}ccccccccc}
  \caption{
 $v$ factors of reference nuclei for which high-$q$ scattering measurements are available.
 They are calculated from the charge distributions whose model is given in the `model' column and on which information can be found in the reference of the `Ref.' column.
 The uncertainty $\sigma_{v}$ is calculated from Eq.~(\ref{eq:unc}) with $q_\mathrm{max}r$.
 The  $v$ factors of elements in which no scattering results are available, I interpolated, as indicated by `Int.' in the `Model' column, with large uncertainties (see Fig.~\ref{fig:nu}).
 \\
  \label{tab:v}
  } 
el. & Z & A & $v\times10^3$ & $\sigma_{v}$ & Model & Ref. & $q_\mathrm{max}\,$fm$^{-1}$ & $q_\mathrm{max} r$\\\\
\endhead

C & 6 & 12 & 2.2  & 0.1 & FB & VJV Ca80 & 4.0 & 1.6  \\
N & 7 & 14 & 2.2  & 0.5 & 3pF & VJV La82 & 2.9 & 1.2  \\
~ & 7 & 15 & 3.3  & 0.5 & FB & Fri  Vr88 & 3.2 & 1.2  \\
O & 8 & 16  & 3.1 & 0.2 & SOG & VJV Si70b & 4.0 & 1.5  \\
F & 9 & 19 & 4.6 & 1.1 & 2pF & VJV Oy75 & 1.8 & 0.6  \\
Ne & 10 & 20 & 2.4 & 1.1 & 3pF & VJV Be85 & 1.8 & 0.6  \\
Na & 11 & 23 & 3.6 & 1.5 & Int. & ~ & ~ & ~  \\
Mg & 12 & 24 & 4.1 & 0.5 & SOG & VJV Li74 & 3.6 & 1.2  \\
~ & 12 & 26 & 3.2 & 0.7 & FB & Fri So88 & 3.0 & 1.0  \\
Al & 13 & 27 & 4.0 & 0.9 & FB & VJV Ro86 & 2.6 & 0.8  \\
Si & 14 & 28 & 5.0 & 0.5 & SOG & VJV Li74 & 3.7 & 1.2  \\
~ & 14 & 29 & 4.2 & 0.9 & FB & VJV Mi82 & 2.6 & 0.8  \\
~ & 14 & 30 & 4.5 & 0.7 & FB & \cite{1997-WE} & 3.0 & 1.0  \\
P & 15 & 31 & 5.3 & 0.8 & FB & \cite{1997-WE} & 2.8 & 0.9  \\
S & 16 & 32 & 5.2 & 0.6 & FB & \cite{1997-WE} & 3.7 & 1.1  \\
~ & 16 & 34 & 5.1 & 0.9 & FB & \cite{1983-Ry83a} & 2.6 & 0.8  \\
~ & 16 & 36 & 4.7 & 0.9 & FB & \cite{1983-Ry83a} & 2.6 & 0.8  \\
Cl & 17 & 35 & 6.9 & 1.2 & 3pF & Br80 & 1.7 & 0.5  \\
~ & 17 & 37 & 5.8 & 1.2 & 3pF & Br80 & 1.7 & 0.5  \\
Ar & 18 & 38 & 5.8 & 1.7 & Int. & ~ & ~ & ~  \\
~  & 18 & 40 & 5.8 & 1.2 & FB & VJV Ot82 & 1.8 & 0.5  \\
K & 19 & 39 & 6.2 & 0.6 & SOG & VJV Li74 & 3.6 & 1.1  \\
Ca & 20 & 40 & 6.9 & 0.6 & SOG & VJV Si79 & 3.7 & 1.1  \\
~ & 20 & 48 & 5.2 & 0.7 & SOG & VJV Si74 & 3.4 & 1.0  \\
Sc & 21 & 45 & 6.6 & 1.7 & Int. & ~ & ~ & ~  \\
Ti & 22 & 48 & 7.2 & 1.1 & FB & VJV Se85 & 2.2 & 0.6  \\
~ & 22 & 50 & 6.7 & 1.1 & FB & VJV Se85 & 2.2 & 0.6  \\
V & 23 & 51 & 6.2 & 1.2 & 2PF & VJV Pe73 & 1.8 & 0.5  \\
Cr & 24 & 50 & 7.5 & 1.0 & FB & VJV Li83c & 2.6 & 0.7  \\
~ & 24 & 52 & 7.4 & 1.0 & FB & VJV Li83c & 2.6 & 0.7  \\
~ & 24 & 54 & 7.2 & 1.0 & FB & VJV Li83c & 2.6 & 0.7  \\
Mn & 25  & 55 & 6.5 & 1.7 & Int. & ~ & ~ & ~  \\
Fe & 26 & 54 & 5.3 & 1.1 & FB & VJV Wo76 & 2.2 & 0.6  \\
~ & 26 & 56 & 5.9 & 1.1 & FB & VJV Wo76 & 2.2 & 0.6  \\
~ & 26 & 58 & 5.7 & 1.1 & FB & VJV Wo76 & 2.2 & 0.6  \\
Co & 27 & 59 & 6.6 & 1.1 & FB & VJV Sc77 & 2.2 & 0.6  \\
Ni & 28 & 58 & 6.6 & 0.7 & SOG & VJV Ca80b & 3.9 & 1.0  \\
Cu & 29 & 63 & 7.3 & 1.1 & FB & VJV Sc77 & 2.2 & 0.6  \\
~ & 29 & 65 & 7.3 & 1.1 & FB & VJV Sc77 & 2.2 & 0.6  \\
Zn & 30 & 64 & 7.0 & 1.1 & FB & WO76 & 2.2 & 0.6  \\
~ & 30 & 66 & 6.7 & 1.1 & FB & WO76 & 2.2 & 0.6  \\
~ & 30 & 68 & 7.1 & 1.1 & FB & WO76 & 2.2 & 0.6  \\
~ & 30 & 70 & 7.6 & 1.1 & FB & WO76 & 2.2 & 0.6  \\
Ga & 31 & 71 & 8.1 & 1.7 & Int. & ~ & ~ & ~  \\
Ge & 32 & 70 & 8.9 & 1.0 & FB & Ma84 & 2.9 & 0.7  \\
~ & 32 & 72 & 8.5 & 1.0 & FB & Ma84 & 2.9 & 0.7  \\
~ & 32 & 74 & 8.0 & 1.0 & FB & Ma84 & 2.9 & 0.7  \\\\
\hline
\end{longtable}

\LTright=0pt
\LTleft=0pt

\clearpage

\normalsize
\begin{longtable}{@{\extracolsep\fill}ccccccccc}

  \caption{
  Reference radii used in this work. Unless stated otherwise in the note, they are determined via Eq.~(\ref{eq:combined}) and~\ref{eq:nu} with the $2P_{3/2}-1S$ Barret radii given in~\cite{2004-Fricke} and the $v$ factors from tab.~\ref{tab:v}.
  Uncertainties are denoted by $\sigma$ and correspond to statistics and energy calibration (exp), nuclear polarization (NP), and charge distribution (CD) as resulting from the $\nu$ factors of Tab.~\ref{tab:v} \\
  \label{tab:rad}
  } 

el. & Z & A & $r_{\mathrm{ch}}$ & $\sigma_{\mathrm{exp}}$ & $\sigma_{\mathrm{NP}}$ & $\sigma_{\mathrm{CD}}$ & $\sigma_{\mathrm{tot}}$ & Note\\\\

\endhead
Li & 3 & 6 & 2.589 & 0.039 & ~ & ~ & 0.039 & A \\ 
Be & 4 & 9 & 2.519 & 0.012 & ~ & 0.030 & 0.032 & B \\  
B  & 5 & 11 & 2.411 & 0.021 & ~ & ~ & 0.021 & C \\  
C  & 6 & 12 & 2.483 & 0.002 & 0.001 & 0.000 & 0.002 & D \\  
N  & 7 & 14 & 2.556 & 0.009 & 0.002 & 0.001  & 0.009 &  \\  
~  & 7 & 15 & 2.612 & 0.009 & ~ & ~ & 0.009   & E \\  
O  & 8 & 16 & 2.701 & 0.004 & 0.001 & 0.001  & 0.004 & F \\  
F  & 9 & 19 & 2.902 & 0.003 & 0.002 & 0.003  & 0.005 & $\dagger$\\  
Ne & 10 & 20 & 3.001 & 0.004 & 0.003 & 0.003 & 0.006 & $\dagger$ \\  
Na & 11 & 23 & 2.992 & 0.002 & 0.002 & 0.005 & 0.006 & $\dagger$ \\  
Mg & 12 & 24 & 3.056 & 0.001 & 0.002 & 0.002 & 0.003 & $\dagger$ \\  
~  & 12 & 26 & 3.030 & 0.001 & 0.002 & 0.002 & 0.003 & $\dagger$ \\  
Al & 13 & 27 & 3.061 & 0.001 & 0.002 & 0.003 & 0.003 & $\dagger$G \\  
Si & 14 & 28 & 3.123 & 0.001 & 0.002 & 0.002 & 0.003 & $\dagger$ \\  
P  & 15 & 31 & 3.190 & 0.001 & 0.002 & 0.002 & 0.003 &  \\  
S  & 16 & 32 & 3.262 & 0.001 & 0.002 & 0.002 & 0.003 &  \\  
~  & 16 & 34 & 3.284 & 0.001 & 0.002 & 0.003 & 0.004 &  \\  
~  & 16 & 36 & 3.298 & 0.001 & 0.001 & 0.003 & 0.003 &  \\  
Cl & 17 & 35 & 3.388 & 0.015 & ~ & ~ & 0.015 & H \\  
Cl & 17 & 37 & 3.384 & 0.015 & ~ & ~ & 0.015 & H \\  
Ar & 18 & 38 & 3.402 & 0.002 & 0.003 & 0.006 & 0.007 &  \\  
~  & 18 & 40 & 3.427 & 0.001 & 0.002 & 0.004 & 0.005 &  \\  
K  & 19 & 39 & 3.435 & 0.001 & 0.001 & 0.002 & 0.003 &  \\  
Ca & 20 & 40 & 3.481 & 0.001 & 0.001 & 0.002 & 0.003 &  \\  
~  & 20 & 48 & 3.475 & 0.001 & 0.001 & 0.003 & 0.003 &  \\  
Sc & 21 & 45 & 3.548 & 0.001 & 0.002 & 0.006 & 0.006 &  \\  
Ti & 22 & 48 & 3.595 & 0.001 & 0.002 & 0.004 & 0.004 & \\  
~  & 22 & 50 & 3.572 & 0.001 & 0.002 & 0.004 & 0.004 & \\  
V  & 23 & 51 & 3.598 & 0.001 & 0.002 & 0.004 & 0.005 &  \\  
Cr & 24 & 50 & 3.664 & 0.000 & 0.002 & 0.004 & 0.004 & \\  
~  & 24 & 52 & 3.644 & 0.000 & 0.002 & 0.004 & 0.004 &  \\  
~  & 24 & 54 & 3.689 & 0.001 & 0.002 & 0.004 & 0.004 &  \\  
Mn & 25 & 55 & 3.705 & 0.001 & 0.002 & 0.006 & 0.007 &  \\  
Fe & 26 & 54 & 3.688 & 0.001 & 0.002 & 0.004 & 0.004 &  I\\  
~  & 26 & 56 & 3.733 & 0.001 & 0.002 & 0.004 & 0.004 &  I\\  
Co & 27 & 59 & 3.786 & 0.001 & 0.002 & 0.004 & 0.005 &  \\  
Ni & 28 & 58 & 3.773 & 0.001 & 0.001 & 0.003 & 0.003 &  J \\  
Cu  & 29 & 65 & 3.902 & 0.0005 & 0.001 & 0.004 & 0.005 &  \\  
Zn  & 30 & 68 & 3.964 & 0.0005 & 0.002 & 0.005 & 0.005 &  K \\  
Ga & 31 & 71 & 4.014 & 0.0001 & 0.001 & 0.007 & 0.007 & \\  
Ge & 32 & 72 & 4.059 & 0.0002 & 0.002 & 0.004 & 0.004 &  \\  
\\
\hline

\end{longtable}
\noindent
\small
A. Estimated through a recent (nearly) model-independent combined analysis of several electron scattering experiments~\cite{2011-Li}.\\
B. Estimated from a model-dependent analysis of an electron scattering experiment covering a narrow momentum transfer range~\cite{1972-Be}. The same publication quotes a radius for $^{12}$C which deviates by 3 standard errors from its modern value. I added a systematic uncertainty to account for this deviation.\\
C. Calculated from the difference $r(^{12}C)-r(^{11}B)=0.072(21)\,$fm measured via elastic pion scattering~\cite{1980-B}.\\
D. Extracted from a combined analysis of high resolution muonic x-ray measurements and electron scattering experiments going beyond the Barret moment recipe~\cite{1982-12C,1984-12C}. This radius agrees with the dispersion-corrected radius from electron scattering~\cite{1991-12C}.
Note that $r(^{12}$C) quoted in~\cite{2013-AM} lies 6 standard deviations away from my recommended value.
This might be due to a propagation of a typo in the uncertainty the Barret radii of carbon isotopes given in~\cite{2004-C}.
\\
E. Based on electron scattering over a wide momentum transfer range~\cite{1988-Nscat}. Dispersion corrections are not included.\\
F. Notice that the statistical uncertainty in the Barret moment given in \cite{2004-OFrick} is overestimated due to a typo.\\
G. An identical radius appears in~\cite{2021-Al}, with a more conservative uncertainty estimate stemming for the charge distribution.\\
H. Muonic x-rays were only measured with a naturally-abundant sample~\cite{1967-Cl}. Radii deduced from electron scattering experiment~\cite{1980-Cl}. Missing uncertainty due to model dependence.\\
I. A different combined analysis method, in which the uncertainties in the electron scattering experiment are accounted for, gives $r_{54}=3.692(5)\,$fm and $r_{56}=3.737(5)\,$fm~\cite{1980-ZnNiFe} after scaling to the modern values of the Barret equivalent radii. \\
J. A different combined analysis method, in which the uncertainties in the electron scattering experiment are accounted for, gives $r_{58}=3.774(5)\,$fm~\cite{1980-ZnNiFe} after scaling to the modern value of the Barret equivalent radius. \\
K. A different combined analysis method, in which the uncertainties in the electron scattering experiment are accounted for, gives $r_{68}=3.966(5)\,$fm~\cite{1980-ZnNiFe} after scaling to the modern value of the Barret equivalent radius.\\
$\dagger$ The daggers denote instances where the experimental uncertainty quoted in Fricke's book differs from that mentioned in the original publication. The former is only statistical, and the latter contains calibration uncertainty as well. For absolute radii, the latter is suitable.

\clearpage
\normalsize
\begin{longtable}{@{\extracolsep\fill}ccccccccc}
  \caption{
Re-calibrated radii of the Neon chain, whose optical isotope shifts are given in \cite{2011-Ne}, using $r_{20}-r_{22}=0.0533(4)\,$fm \cite{2022-Sailer}, $F_{214}=-30.5(1.5)\,$MHz/fm$^2$ \cite{2021-NeF,2019-Ne}, and $r_{20}=3.001(6)\,$fm from Tab. \ref{tab:rad}.\\
  \label{tab:Ne}
  } 
A & $\delta r^2_{20,A}$ & $\sigma_{\mathrm{stat}}$ & $\sigma_{\mathrm{K,F}}$ & $\sigma_{\mathrm{tot}}$ & $r_A$ & $\sigma_{\mathrm{\delta r^2}}$ & $\sigma_{\mathrm{r_{20}}}$ & $ \sigma_{\mathrm{tot}}$ \\\\
\endhead

17 & 0.082   & 0.038 & 0.030 & 0.049 & 3.015 & 0.008 & 0.006 & 0.010 \\
18 & -0.401  & 0.020 & 0.042 & 0.047 & 2.934 & 0.008 & 0.006 & 0.010 \\
19 & -0.038  & 0.024 & 0.012 & 0.027 & 2.995 & 0.005 & 0.006 & 0.007 \\
20 & 0       &     ~ &     ~ &     ~ & 3.001 & ~ & 0.006 & 0.006 \\
21 & -0.230  & 0.018 & 0.005 & 0.019 & 2.963 & 0.003 & 0.006 & 0.007 \\
22 & -0.317 & ~     & ~ & 0.002 & 2.948 & 0.000 & 0.006 & 0.006 \\
23 & -0.601  & 0.045 & 0.012 & 0.047 & 2.899 & 0.008 & 0.006 & 0.010 \\
24 & -0.634  & 0.025 & 0.011 & 0.027 & 2.894 & 0.005 & 0.006 & 0.008 \\
25 & -0.336  & 0.022 & 0.023 & 0.032 & 2.945 & 0.005 & 0.006 & 0.008 \\
26 & -0.374  & 0.023 & 0.028 & 0.036 & 2.938 & 0.006 & 0.006 & 0.009 \\
28 & 0.006   & 0.046 & 0.056 & 0.073 & 3.002 & 0.012 & 0.006 & 0.013 \\\\
\hline
 
\end{longtable}

\LTright=0pt
\LTleft=0pt


\normalsize
\begin{longtable}{@{\extracolsep\fill}cclc cclc cccl cc}
  \caption{
Input mirror shifts $\Delta_I$ to the mirror fit, based on the reference radii of Tab.~\ref{tab:rad}, with radii differences taken from the references given in Tab.~\ref{tab:mirr}\\
  \label{tab:fit}
  } 
$A$  & $I$    & el. & $Z$ & $N$ & $r\,$fm & el. & $Z$ & $N$ & $r\,$fm  & $\Delta_I\,$fm  & $\Delta_I$/I~fm & Weight fm$^{-2}$ & n$\sigma$\\\\
\endhead
7  & 0.14 & Li & 3  & 4  & 2.449(41) & Be & 4  & 3  & 2.646(33) & 0.197(53) & 1.42(37) & 7   & 0.0 \\
18 & 0.11 & O  & 8  & 10 & 2.777(07) & Ne & 10 & 8  & 2.934(09) & 0.161(15) & 1.45(13) & 96  & 0.3 \\
19 & 0.05 & F  & 9  & 10 & 2.902(04) & Ne & 10 & 9  & 2.995(06) & 0.093(07) & 1.77(14) & 52  & 2.7 \\
22 & 0.09 & Ne & 10 & 12 & 2.948(04) & Mg & 12 & 10 & 3.071(05) & 0.123(07) & 1.35(07) & 186 & 0.5 \\
23 & 0.04 & Na & 11 & 12 & 2.992(05) & Mg & 12 & 11 & 3.043(04) & 0.051(07) & 1.17(15) & 43  & 1.4 \\
32 & 0.13 & Si & 14 & 18 & 3.154(13) & Ar & 18 & 14 & 3.346(17) & 0.193(21) & 1.54(17) & 36  & 0.9 \\
34 & 0.06 & S  & 16 & 18 & 3.284(04) & Ar & 18 & 16 & 3.365(11) & 0.081(12) & 1.37(20) & 24  & 0.1 \\
36 & 0.11 & S  & 16 & 20 & 3.298(04) & Ca & 20 & 16 & 3.452(06) & 0.155(07) & 1.39(06) & 270 & 0.0 \\
37 & 0.03 & Ar & 18 & 19 & 3.390(07) & K  & 19 & 18 & 3.419(10) & 0.029(13) & 1.09(47) & 5   & 0.7 \\
38 & 0.05 & Ar & 18 & 20 & 3.402(06) & Ca & 20 & 18 & 3.469(03) & 0.067(08) & 1.28(15) & 47  & 0.8 \\
39 & 0.03 & K  & 19 & 20 & 3.435(04) & Ca & 20 & 19 & 3.464(04) & 0.029(06) & 1.14(22) & 21  & 1.2 \\
54 & 0.04 & Fe & 26 & 28 & 3.688(04) & Ni & 28 & 26 & 3.741(05) & 0.053(06) & 1.44(17) & 34  & 0.3 \\\\
\hline
\end{longtable}

\clearpage
\normalsize
\begin{longtable}{@{\extracolsep\fill}cclcccllcccl}
  \caption{
Charge radii of mirror nuclei. Known radii are taken from Tab. \ref{tab:rad} along with isotopic differences taken from the indicated references. Predicted radii relying on the validity of Eq.~(\ref{eq:fit}) are given in italics.\\
  \label{tab:mirr}
  } 
        A & I & el. & Z & N & r & Ref./Note & el. & Z & N & $r$  & Ref./Note \vspace{2 mm}\\
\endhead

$7 $ & $0.14$ & Be & $4 $ & $3 $ & $2.646(33)$ & \cite{2012-Be}                    & Li & $3 $ & $4 $ & $2.449(41)$ & \cite{2011-LiComb, 2013-Brown} \\
$8 $ & $0.25$ & B  & $5 $ & $3 $ & $\mathit{2.685(45)}$ &                           & Li & $3 $ & $5 $ & $2.339(44)$ & \cite{2011-LiPRL, 2011-Li} \\
$9 $ & $0.33$ & C  & $6 $ & $3 $ & $\mathit{2.705(47)}$ &                           & Li & $3 $ & $6 $ & $2.245(46)$ & \cite{2011-LiPRL, 2011-Li} \\
$10$ & $0.20$ & C  & $6 $ & $4 $ & $\mathit{2.638(36)}$ &                           & Be & $4 $ & $6 $ & $2.361(36)$ & \cite{2012-Be}\\
$11$ & $0.09$ & C  & $6 $ & $5 $ & $\mathit{2.536(21)}$ &                           & B  & $5 $ & $6 $ & $2.411(21)$ & ~ \\
$13$ & $0.08$ & N  & $7 $ & $6 $ & $\mathit{2.564(04)}$ &                           & C  & $6 $ & $7 $ & $2.458(02)$ & \cite{2024-13C} \\
$14$ & $0.14$ & O  & $8 $ & $6 $ & $\mathit{2.705(10)}$ &                           & C  & $6 $ & $8 $ & $2.508(09)$ & \cite{1982-14C} A\\
$15$ & $0.07$ & O  & $8 $ & $7 $ & $\mathit{2.704(09)}$ &                           & N  & $7 $ & $8 $ & $2.612(09)$ & ~ \\
$17$ & $0.06$ & F  & $9 $ & $8 $ & $\mathit{2.774(08)}$ &                           & O  & $8 $ & $9 $ & $2.693(08)$ & \cite{1979-17O} B\\
$17$ & $0.18$ & Ne & $10$ & $7 $ & $3.015(10)$          & Tab. \ref{tab:Ne}         & N  & $7 $ & $10$ & $\mathit{2.771(12)}$ & \\
$18$ & $0.11$ & Ne & $10$ & $8 $ & $2.934(10)$          & Tab. \ref{tab:Ne}         & O  & $8 $ & $10$ & $2.777(07)$ & \cite{1980-O} C\\
$19$ & $0.05$ & Ne & $10$ & $9 $ & $2.995(07)$          & Tab. \ref{tab:Ne}         & F  & $9 $ & $10$ & $2.902(05)$ & ~ \\
$20$ & $0.10$ & Na & $11$ & $9 $ & $\mathit{2.983(25)}$ & Tab. \ref{tab:Na}         & F  & $9 $ & $11$ & $\mathit{2.845(25)}$ &  \\
$21$ & $0.05$ & Na & $11$ & $10$ & $\mathit{3.029(07)}$ &                           & Ne & $10$ & $11$ & $2.963(07)$ & Tab. \ref{tab:Ne}  \\
$21$ & $0.14$ & Mg & $12$ & $9 $ & $3.067(07)$          & \cite{2012-Mg,2022-NaMg}  & F  & $9 $ & $12$ & $\mathit{2.869(09)}$  \\
$22$ & $0.09$ & Mg & $12$ & $10$ & $3.071(05)$          & \cite{2012-Mg,2022-NaMg}  & Ne & $10$ & $12$ & $2.948(06)$ & \cite{2022-Sailer} \\
$23$ & $0.04$ & Mg & $12$ & $11$ & $3.043(04)$          & \cite{2012-Mg,2022-NaMg}  & Na & $11$ & $12$ & $2.992(06)$ & ~ \\
$23$ & $0.13$ & Al & $13$ & $10$ & $\mathit{3.080(11)}$ &                           & Ne & $10$ & $13$ & $2.899(10)$ & Tab. \ref{tab:Ne} \\
$24$ & $0.08$ & Al & $13$ & $11$ & $\mathit{3.078(11)}$ &                           & Na & $11$ & $13$ & $\mathit{2.963(11)}$ & Tab. \ref{tab:Na} \\
$24$ & $0.17$ & Si & $14$ & $10$ & $\mathit{3.124(10)}$ &                           & Ne & $10$ & $14$ & $2.894(08)$ & Tab. \ref{tab:Ne} \\
$25$ & $0.04$ & Al & $13$ & $12$ & $\mathit{3.082(04)}$ &                           & Mg & $12$ & $13$ & $3.026(03)$ & \cite{2012-Mg,2022-NaMg} \\
$25$ & $0.12$ & Si & $14$ & $11$ & $\mathit{3.129(15)}$ &                           & Na & $11$ & $14$ & $\mathit{2.963(14)}$ & Tab. \ref{tab:Na} \\
$26$ & $0.08$ & Si & $14$ & $12$ & $\mathit{3.136(04)}$ &                           & Mg & $12$ & $14$ & $3.030(03)$ & ~ \\
$26$ & $0.15$ & P  & $15$ & $11$ & $\mathit{3.190(17)}$ &                           & Na & $11$ & $15$ & $\mathit{2.977(16)}$ & Tab. \ref{tab:Na} \\
$27$ & $0.04$ & Si & $14$ & $13$ & $\mathit{3.111(04)}$ &                           & Al & $13$ & $14$ & $3.060(03)$ & ~ \\
$27$ & $0.11$ & P  & $15$ & $12$ & $\mathit{3.181(05)}$ &                           & Mg & $12$ & $15$ & $3.027(03)$ & \cite{2012-Mg,2022-NaMg} \\
$27$ & $0.19$ & S  & $16$ & $11$ & $\mathit{3.256(22)}$ &                           & Na & $11$ & $16$ & $\mathit{3.000(21)}$ & Tab. \ref{tab:Na} \\
$28$ & $0.07$ & P  & $15$ & $13$ & $\mathit{3.157(09)}$ &                           & Al & $13$ & $15$ & $3.058(09)$ & \cite{2021-Al,2024-Al} \\
$28$ & $0.14$ & S  & $16$ & $12$ & $\mathit{3.261(06)}$ &                           & Mg & $12$ & $16$ & $3.063(04)$ & \cite{2012-Mg,2022-NaMg} \\
$29$ & $0.03$ & P  & $15$ & $14$ & $\mathit{3.165(06)}$ &                           & Si & $14$ & $15$ & $3.118(06)$ & \cite{2023-Si} \\
$29$ & $0.10$ & S  & $16$ & $13$ & $\mathit{3.221(15)}$ &                           & Al & $13$ & $16$ & $3.078(15)$ & \cite{2021-Al,2024-Al} \\
$30$ & $0.07$ & S  & $16$ & $14$ & $\mathit{3.226(06)}$ &                           & Si & $14$ & $16$ & $3.134(05)$ & \cite{2023-Si} \\
$31$ & $0.03$ & S  & $16$ & $15$ & $\mathit{3.235(03)}$ &                           & P  & $15$ & $16$ & $3.190(03)$ & ~ \\
$31$ & $0.16$ & Ar & $18$ & $13$ & $\mathit{3.323(31)}$ &                           & Al & $13$ & $18$ & $3.099(31)$ & \cite{2021-Al,2024-Al} \\
$32$ & $0.13$ & Ar & $18$ & $14$ & $3.346(17)$          & \cite{1996-Ar,2008-Ar} D & Si     & $14$ & $18$ & $3.154(13)$ & \cite{2023-Si}  \\
$33$ & $0.09$ & Ar & $18$ & $15$ & $3.343(15)$          & \cite{1996-Ar,2008-Ar} D & P      & $15$ & $18$ & $\mathit{3.217(15)}$  \\
$34$ & $0.06$ & Ar & $18$ & $16$ & $3.365(11)$          & \cite{1996-Ar,2008-Ar} D & S      & $16$ & $18$ & $3.284(04)$ & ~ \\
$35$ & $0.03$ & Ar & $18$ & $17$ & $3.363(11)$          & \cite{1996-Ar,2008-Ar} D & Cl     & $17$ & $18$ & $\mathit{3.323(11)}$ \\
$36$ & $0.06$ & K  & $19$ & $17$ & $3.406(14)$          & \cite{2015-36-37K,2021-K}   & Cl     & $17$ & $19$ & $\mathit{3.329(14)}$ \\
$36$ & $0.11$ & Ca & $20$ & $16$ & $3.452(06)$          & \cite{2019-36-38Ca,2020-Kra}     & S      & $16$ & $20$ & $3.298(04)$\\
$37$ & $0.03$ & K  & $19$ & $18$ & $3.419(10)$          & \cite{2015-36-37K,2021-K}   & Ar     & $18$ & $19$ & $3.390(07)$ & \cite{2008-Ar}  \\
$37$ & $0.08$ & Ca & $20$ & $17$ & $3.451(06)$          & \cite{2019-36-38Ca,2020-Kra}     & Cl     & $17$ & $20$ & $\mathit{3.338(07)}$ \\
$38$ & $0.05$ & Ca & $20$ & $18$ & $3.469(04)$          & \cite{2019-36-38Ca,2020-Kra}     & Ar     & $18$ & $20$ & $3.402(06)$  \\
$39$ & $0.03$ & Ca & $20$ & $19$ & $3.464(04)$          & \cite{2019-36-38Ca,2020-Kra}     & K      & $19$ & $20$ & $3.435(04)$ & ~ \\
$40$ & $0.05$ & Sc & $21$ & $19$ & $\mathit{3.508(04)}$ &                         & K  & $19$ & $21$ & $3.439(04)$ & \cite{2021-K,2006-40-41K,2008-40K} \\
$40$ & $0.10$ & Ti & $22$ & $18$ & $\mathit{3.566(06)}$ &                         & Ar & $18$ & $22$ & $3.427(04)$   \\
$41$ & $0.02$ & Sc & $21$ & $20$ & $\mathit{3.515(04)}$ &                         & Ca & $20$ & $21$ & $3.481(04)$ & \cite{2020-Kra,1984-Ca} \\
$41$ & $0.07$ & Ti & $22$ & $19$ & $\mathit{3.556(06)}$ &                         & K  & $19$ & $22$ & $3.455(05)$ & \cite{2021-K,2006-40-41K,2008-40K,1982-K} \\
$42$ & $0.05$ & Ti & $22$ & $20$ & $\mathit{3.576(05)}$ &                         & Ca & $20$ & $22$ & $3.510(04)$ & \cite{2020-Ca} \\
$42$ & $0.14$ & Cr & $24$ & $18$ & $\mathit{3.638(12)}$ &                         & Ar & $18$ & $24$ & $3.440(11)$ & \cite{2008-Ar} D \\
$43$ & $0.02$ & Ti & $22$ & $21$ & $\mathit{3.583(12)}$ &                         & Sc & $21$ & $22$ & $3.550(11)$ & \cite{2011-42-46Sc,2023-Sc} E\\
$43$ & $0.07$ & V  & $23$ & $20$ & $\mathit{3.594(05)}$ &                         & Ca & $20$ & $23$ & $3.497(04)$ & \cite{2020-Ca} \\
$43$ & $0.12$ & Cr & $24$ & $19$ & $\mathit{3.620(09)}$ &                         & K  & $19$ & $24$ & $3.459(08)$ & \cite{2021-K,1982-K} \\
$44$ & $0.05$ & V  & $23$ & $21$ & $\mathit{3.604(08)}$ &                         & Sc & $21$ & $23$ & $3.540(08)$ & \cite{2011-42-46Sc,2023-Sc} E\\
$44$ & $0.09$ & Cr & $24$ & $20$ & $\mathit{3.645(06)}$ &                         & Ca & $20$ & $24$ & $3.519(04)$ & \cite{2020-Ca} \\
$45$ & $0.02$ & V  & $23$ & $22$ & $\mathit{3.628(05)}$ &                         & Ti & $22$ & $23$ & $3.597(05)$ & \cite{2004-Ti} \\
$45$ & $0.07$ & Cr & $24$ & $21$ & $\mathit{3.640(07)}$ &                         & Sc & $21$ & $24$ & $3.548(07)$ & ~ \\
$46$ & $0.04$ & Cr & $24$ & $22$ & $\mathit{3.670(05)}$ &                         & Ti & $22$ & $24$ & $3.610(04)$ & \cite{2004-Ti} \\
$46$ & $0.09$ & Mn & $25$ & $21$ & $\mathit{3.651(09)}$ &                         & Sc & $21$ & $25$ & $3.530(08)$ & \cite{2011-42-46Sc,2023-Sc} E\\
$46$ & $0.13$ & Fe & $26$ & $20$ & $\mathit{3.677(07)}$ &                         & Ca & $20$ & $26$ & $3.497(04)$ & \cite{2020-Ca} \\
$47$ & $0.06$ & Mn & $25$ & $22$ & $\mathit{3.688(05)}$ &                         & Ti & $22$ & $25$ & $3.599(04)$ & \cite{2004-Ti} \\
$48$ & $0.08$ & Fe & $26$ & $22$ & $\mathit{3.710(05)}$ &                         & Ti & $22$ & $26$ & $3.595(04)$ & ~ \\
$49$ & $0.02$ & Mn & $25$ & $24$ & $\mathit{3.685(06)}$ &                         & Cr & $24$ & $25$ & $3.656(06)$ & \cite{2023-Cr} \\
$50$ & $0.04$ & Fe & $26$ & $24$ & $\mathit{3.709(04)}$ &                         & Cr & $24$ & $26$ & $3.664(04)$ &  \\
$50$ & $0.12$ & Ni & $28$ & $22$ & $\mathit{3.738(06)}$ &                         & Ti & $22$ & $28$ & $3.572(04)$ & ~ \\
$51$ & $0.02$ & Fe & $26$ & $25$ & $\mathit{3.746(35)}$ &                         & Mn & $25$ & $26$ & $3.719(35)$ & \cite{2010-Mn,2016-Mn} F\\
$51$ & $0.06$ & Co & $27$ & $24$ & $\mathit{3.718(05)}$ &                         & Cr & $24$ & $27$ & $3.637(04)$ & \cite{2023-Cr}  \\
$51$ & $0.10$ & Ni & $28$ & $23$ & $\mathit{3.734(06)}$ &                         & V  & $23$ & $28$ & $3.598(04)$ & ~ \\
$52$ & $0.04$ & Co & $27$ & $25$ & $\mathit{3.728(27)}$ &                         & Mn & $25$ & $27$ & $3.675(27)$ & \cite{2010-Mn,2016-Mn} F\\
$52$ & $0.08$ & Ni & $28$ & $24$ & $\mathit{3.750(05)}$ &                         & Cr & $24$ & $28$ & $3.644(04)$ & ~ \\
$53$ & $0.02$ & Co & $27$ & $26$ & $\mathit{3.727(07)}$ &                         & Fe & $26$ & $27$ & $3.701(07)$ & \cite{2016-53Fe} \\
$53$ & $0.06$ & Ni & $28$ & $25$ & $\mathit{3.745(22)}$ &                         & Mn & $25$ & $28$ & $3.667(22)$ & \cite{2010-Mn,2016-Mn} F\\
$54$ & $0.04$ & Ni & $28$ & $26$ & $3.741(05)$          & \cite{2021-54Ni}        & Fe & $26$ & $28$ & $3.688(04)$ & ~ \\
$55$ & $0.02$ & Ni & $28$ & $27$ & $3.691(05)$          & \cite{2022-5556Ni}      & Co & $27$ & $28$ & $\mathit{3.666(05)}$  \\
$56$ & $0.07$ & Zn & $30$ & $26$ & $\mathit{3.832(05)}$ &                         & Fe & $26$ & $30$ & $3.733(04)$ & ~ \\
$58$ & $0.03$ & Zn & $30$ & $28$ & $\mathit{3.820(03)}$ &                         & Ni & $28$ & $30$ & $3.773(03)$ & ~ \\
$59$ & $0.02$ & Zn & $30$ & $29$ & $\mathit{3.844(10)}$ &                         & Cu & $29$ & $30$ & $3.820(10)$ & \cite{2016-Cu} \\
$60$ & $0.03$ & Ga & $31$ & $29$ & $\mathit{3.883(09)}$ &                         & Cu & $29$ & $31$ & $3.836(09)$ & \cite{2016-Cu} \\
$60$ & $0.07$ & Ge & $32$ & $28$ & $\mathit{3.903(05)}$ &                         & Ni & $28$ & $32$ & $3.809(03)$ & \cite{2022-5556Ni}   \\
$61$ & $0.05$ & Ge & $32$ & $29$ & $\mathit{3.924(07)}$ &                         & Cu & $29$ & $32$ & $3.856(07)$ & \cite{2016-Cu} \\
$62$ & $0.03$ & Ge & $32$ & $30$ & $\mathit{3.927(06)}$ &                         & Zn & $30$ & $32$ & $3.883(06)$ & \cite{2019-Zn,2023-Zn} \\
$63$ & $0.02$ & Ge & $32$ & $31$ & $\mathit{3.955(19)}$ &                         & Ga & $31$ & $32$ & $3.933(19)$ & \cite{2012-Ga} \\
$64$ & $0.03$ & As & $33$ & $31$ & $\mathit{3.984(17)}$ &                         & Ga & $31$ & $33$ & $3.941(17)$ & \cite{2012-Ga} \\
$65$ & $0.05$ & Se & $34$ & $31$ & $\mathit{4.025(14)}$ &                         & Ga & $31$ & $34$ & $3.961(14)$ & \cite{2017-65Ga} \\
$74$ & $0.03$ & Sr & $38$ & $36$ & $\mathit{4.205(12)}$ &                         & Kr & $36$ & $38$ & $4.168(12)$ & \cite{1995-Kr,2004-KrFrick} G\vspace{2 mm}\\
 \hline
\end{longtable}
\noindent
\small
A. The absolute radius of $^{14}$C was measured via muonic x-ray spectroscopy to be $2.496(19)\,$fm~\cite{1982-14C}. It was quoted later with half of that uncertainty \cite{1998-Ang} which persisted through later compilations \cite{2004-Ang, 2013-AM}.
The radius given in this work is obtained by combining that of $^{12}$C from \cite{1984-12C} together with the radii difference from \cite{1982-14C}, for which the calibration uncertainty is reduced.\\
B. The difference $r_{16}-r_{17}=8(7)\,$am was measured with electron scattering over a wide momentum transfer range analyzed (nearly) model-independently~\cite{1979-17O}. The following note indicates the reliability of this value.\\
C. The difference measured via muonic x-rays $r_{18}-r_{16}=76(5)\,$am \cite{1980-O} agrees well with that measured with electron scattering $r_{18}-r_{16}=74(5)\,$am~\cite{1979-17O}. \\
D. The uncertainty in these radii is dominated by a $10\,$\% uncertainty assumed for a semi-empirically evaluated field shift factor. Experience shows that this uncertainty might be underestimated. See e.g.~\cite{1985-FS, 1992-Ca, 2006-Mg, 2019-Ne, 2024-Ag}. \\
E. The optical isotope shift measurements of \cite{2011-42-46Sc} were reanalyzed in~\cite{2023-Sc} using more reliable IS factor calculations.\\
F. Systematic uncertainties related to the atomic factors, missing in a recent compilation~\cite{2021-Wang}, were extracted from Fig. 9 in~\cite{2016-Mn} and added in quadrature.\\
%
%
G. The optical measurements of \cite{1995-Kr} were originally analyzed using a semi-empirical field shift factor assuming it has an uncertainty of $10\,\%$, which I believe to be underestimated (See note D).
The extracted radius is given in \cite{2013-AM} without the dominating systematic uncertainty.
I opted for the analysis of \cite{2004-KrFrick} in which the atomic factors are extracted from a King plot.\\
%


\normalsize
\begin{longtable}{@{\extracolsep\fill}cccccccccc}
  \caption{
Re-calibration of the Sodium chain, whose optical isotope shifts are given in \cite{1975-NaIS, 1977-NaIS, 1978-NaIS, 1982-NaIS}, using $F=-39.3(3)\,$MHz/fm$^2$~\cite{2022-NaMg} and $K=384.7(6)\,$MHz/fm$^2$ whose estimation is described in the main text.\label{tab:Na}\vspace{2 mm}} 
 A & $\delta r^2$ & $\sigma_\mathrm{exp}$ & $\sigma_{K, F}$ & $\sigma_\mathrm{tot}$ & r & $\sigma_{\delta r}$ & $\sigma_{r23}$ & $\sigma_\mathrm{tot}$ & Ref.~\cite{2022-NaMg} \\
 ~ & fm$^2$ & fm$^2$ & fm$^2$ & fm$^2$ & fm & fm & fm & fm & fm\vspace{2 mm}
 \endhead
20 & -0.05 & 0.09 & 0.10 & 0.14 & 2.983 & 0.023 & 0.005 & 0.024 & 2.89(9) ~ ~\\
21 & 0.22  &      & ~    & 0.05 & 3.029 & ~     & ~     & 0.006 & 2.97(6) ~ ~\\
22 & 0.01  & 0.02 & 0.03 & 0.04 & 2.994 & 0.006 & 0.005 & 0.008 & 2.967(27)  \\
23 &       &      &    ~ & ~    & 2.992 & ~     & ~     & 0.005 & 2.994(4)~ \\
24 & -0.17 & 0.04 & 0.03 & 0.05 & 2.963 & 0.009 & 0.005 & 0.010 & 2.990(26)\\
25 & -0.18 & 0.05 & 0.05 & 0.07 & 2.963 & 0.012 & 0.005 & 0.013 & 3.013(48)\\
26 & -0.09 & 0.02 & 0.08 & 0.08 & 2.977 & 0.014 & 0.005 & 0.015 & 3.049(68)\\
27 & 0.05  & 0.05 & 0.10 & 0.11 & 3.000 & 0.019 & 0.005 & 0.020 & 3.091(86)\\
28 & 0.23  & 0.07 & 0.12 & 0.14 & 3.030 & 0.023 & 0.005 & 0.024 & 3.14(10) ~\\
29 & 0.62  & 0.10 & 0.14 & 0.17 & 3.094 & 0.028 & 0.005 & 0.028 & 3.22(11) ~\\
30 & 0.81  & 0.15 & 0.16 & 0.22 & 3.125 & 0.035 & 0.005 & 0.036 & 3.26(13) ~\\
31 & 1.23  & 0.09 & 0.18 & 0.20 & 3.192 & 0.031 & 0.005 & 0.031 & 3.34(14)~\vspace{2 mm}\\
\hline

\end{longtable}

\clearpage

\LTright=0pt
\LTleft=0pt
\normalsize
\begin{longtable}{@{\extracolsep\fill}llllllllll}
  \caption{Radii of isotriplet nuclei. $r_{\pm1}$ are from Tab.~\ref{tab:mirr}, $r_{0,\mathrm{SE}}$, $\Delta M_B^{(1)}$, and $r_\mathrm{CW}$ are calculated using Eq.~(\ref{eq:r0}), Eq.~(\ref{eq:DeltaM}), and~(\ref{eq:rwk}), respectively. Asterisks denote short-lived excited nuclear states.``m" denotes long-lived nuclear isomers. $r_{0,\mathrm{exp}}$ are determined by combining reference radii from Tab.~\ref{tab:rad}, optical isotopes shifts given in Refs.~\cite{2023-Al, 2014-38mK, 2011-42-46Sc, 2016-Mn, 2011-Rb}, and improved calculations of atomic factors from Refs.~\cite{2021-K, 2023-Sc, 2024-Al}. \label{tab:iso}\vspace{2 mm}}
  & \multicolumn{1}{l}{$r_{-1}$~fm} & \multicolumn{2}{c}{$r_{0,\mathrm{SE}}$~fm} & $r_{0,\mathrm{exp}}$ fm & \multicolumn{2}{c}{$r_{+1}$~fm}  & $\Delta M_B^{(1)}~$fm$^2$ & $r^2_\mathrm{CW}$ fm$^2$ & Ref.~\cite{2022-Seng}  \\
\endhead
  
~ $^{10}_6$C     & $2.638(36)$ & ~ $^{10}_5$B*     & $2.531(38)$ &             & ~ $^{10}_4$Be     & $2.361(36)$ & $ $           & ~ $9.72(25)$ & N/A         \\
~ $^{14}_8 $O    & $2.705(10)$ & ~ $^{14}_7$N*     & $2.623(09)$ &             & ~ $^{14}_6$C      & $2.508(09)$ & $ $           & $10.41(12)$ & N/A         \\
~ $^{18}_{10}$Ne & $2.934(10)$ & ~ $^{18}_9$F*     & $2.861(07)$ &             & ~ $^{18}_8$O      & $2.777(07)$ & $ $           & $12.08(12)$ & $13.4(5)$   \\
~ $^{22}_{12}$Mg & $3.071(05)$ & ~ $^{22}_{11}$Na* & $3.017(05)$ &             & ~ $^{22}_{10}$Ne  & $2.948(04)$ & $ $           & $13.24(12)$ & $12.9(7)$   \\
~ $^{26}_{14}$Si & $3.136(04)$ & ~ $^{26m}_{13}$Al & $3.088(04)$ & $3.132(08)$ & ~ $^{26}_{12}$Mg  & $3.030(03)$ & $-3.5(0.7)$   & $13.77(12)$ & N/A         \\
~ $^{30}_{16}$S  & $3.226(06)$ & ~ $^{30}_{15}$P*  & $3.181(06)$ &             & ~ $^{30}_{14}$Si  & $3.132(06)$ & $ $           & $14.50(13)$ & N/A         \\
~ $^{34}_{18}$Ar & $3.365(11)$ & ~ $^{34}_{17}$Cl  & $3.327(04)$ &             & ~ $^{34}_{16}$S   & $3.284(04)$ & $ $           & $15.66(13)$ & $15.6(5)$   \\
~ $^{38}_{20}$Ca & $3.469(03)$ & ~ $^{38m}_{19}$K  & $3.440(06)$ & $3.437(05)$ & ~ $^{38}_{18}$Ar  & $3.402(06)$ & $-0.3(0.7)$  & $16.53(13)$ & $16.0(3)$   \\
~ $^{42}_{22}$Ti & $3.576(03)$ & ~ $^{42}_{21}$Sc  & $3.545(04)$ & $3.558(16)$ & ~ $^{42}_{20}$Ca  & $3.510(04)$ & $-2.0(2.4)$   & $17.46(13)$ & $21.5(3.6)$ \\  
~ $^{46}_{24}$Cr & $3.670(04)$ & ~ $^{46}_{23}$V   & $3.642(05)$ &             & ~ $^{46}_{22}$Ti  & $3.610(04)$ & $ $           & $18.29(14)$ & N/A          \\
~ $^{50}_{26}$Fe & $3.709(04)$ & ~ $^{50}_{25}$Mn  & $3.692(04)$ & $3.728(41)$ & ~ $^{50}_{24}$Cr  & $3.664(04)$ & $-6.6(7.8)$   & $18.73(14)$ & $23.2(3.8)$ \\
~ $^{54}_{28}$Ni & $3.741(05)$ & ~ $^{54}_{27}$Co  & $3.715(04)$ &             & ~ $^{54}_{26}$Fe  & $3.688(04)$ & $ $           & $18.93(14)$ & $18.3(9)$   \\
~ $^{58}_{30}$Zn & $3.820(03)$ & ~ $^{58}_{29}$Cu* & $3.797(03)$ &             & ~ $^{58}_{28}$Ni  & $3.773(03)$ & $ $           & $19.66(14)$ & N/A         \\
~ $^{62}_{32}$Ge & $3.927(07)$ & ~ $^{62}_{31}$Ga  & $3.906(06)$ &             & ~ $^{62}_{30}$Zn  & $3.883(06)$ & $ $           & $20.65(15)$ & N/A         \\
~ $^{74}_{38}$Sr & $4.205(12)$ & ~ $^{74}_{37}$Rb  & $4.187(12)$ & $4.194(17)$ & ~ $^{74}_{36}$Kr  & $4.168(12)$ & $-2.0(6.5)$   & $23.32(19)$ & $19.5(5.5)$ \\\\
\hline
 \end{longtable}

\clearpage
\bibliographystyle{unsrtnat}

\bibliography{sample}

 \end{document}